\documentclass[
pre,
aps,
a4paper,
english,
reprint,
twocolumn,
superscriptaddress
]{revtex4}

\usepackage{multirow}
\usepackage{makecell}
\usepackage{array}
\usepackage{mathtools}
\usepackage[dvipsnames]{xcolor}
\usepackage{lipsum}
\usepackage{tikz}
\usepackage{wrapfig}
\usepackage[T1]{fontenc}
\usepackage[utf8]{inputenc}
\usepackage{amsmath}
\usepackage{amssymb}
\usepackage{graphicx}
\usepackage{ctable}
\usepackage{multirow}
\usepackage{blkarray,booktabs,bigstrut}

\usepackage[
citecolor=blue,
colorlinks,
linkcolor=blue,
urlcolor=blue,
]{hyperref}
\usepackage{dcolumn}
\usepackage{bm}
\usepackage{enumerate}
\usepackage{tabularx,booktabs}
\usepackage{multirow}
\usepackage{float}
\usepackage{array}
\usepackage{tabularx}
\usepackage{multirow}
\begin{document}
	
\preprint{AIP/123-QED}
	
\title{{Replicator-mutator dynamics of Rock-Paper-Scissors game: Learning through mistakes}}
\author{Suman Chakraborty}
\email{suman@iitk.ac.in}
\affiliation{
	Department of Physics,
	Indian Institute of Technology Kanpur,
	Uttar Pradesh 208016, India
}
\author{Ishita Agarwal}
\email{ishitaa20@iitk.ac.in}
\affiliation{
	Department of Physics,
	Indian Institute of Technology Kanpur,
	Uttar Pradesh 208016, India
}
\author{Sagar Chakraborty}
\email{sagarc@iitk.ac.in}
\affiliation{
	Department of Physics,
	Indian Institute of Technology Kanpur,
	Uttar Pradesh 208016, India
}
\begin{abstract}
We generalize the Bush--Mosteller learning, the Roth--Erev learning, and the social learning to include mistakes such that the nonlinear replicator-mutator equation with either additive or multiplicative mutation is generated in an asymptotic limit. Subsequently, we exhaustively investigate the ubiquitous Rock-Paper-Scissors game for some analytically tractable motifs of mutation pattern. We consider both symmetric and asymmetric game interactions, and reveal that mistakes can sometimes help the players learn. While the replicator-mutator flow exhibits rich dynamics that include limit cycles and chaotic orbits, it can control chaos as well to lead to rational Nash equilibrium outcome. Moreover, we also report an instance of hitherto unknown Hamiltonian structure of the replicator-mutator equation.
\end{abstract}
	\date{\today}
	\maketitle
\section{Introduction}
It is well-known that while attending to a task, actions that lead to positive outcome are repeated more often (law of effect~\cite{gray2010psychology}) and the time taken to accomplish the task reduces over repeated trials (law of practice~\cite{snoddy1926jap}): In other words, the doer \emph{learns}. Learning is important not only for human beings but also for animals~{\cite{cheney2018book}}, insects~\cite{Dong2023science}, birds~{\cite{charrier2005BP}}, bacteria~{\cite{Taga2003PNAS}}, and other living organisms. In a game-theoretic setup where two players interact, the complete learning process involves for each player a choice of strategy---based on a belief about the opponent---that leads to some outcome that in turn leads to updated belief (through a learning rule~\cite{friedman2016oup}); the strategy adopted in the subsequent round of play is based, via a decision rule, on the updated belief, and this cycle repeats until the optimal strategy is learnt. When the step of belief updating is bypassed, i.e., when the strategy update is directly based on the outcome, the learning process is termed as reinforcement learning. Some of the well-known reinforcement learning mechanisms---like the Roth--Erev learning~\cite{Roth1995GEB,erev1998JSTOR}, the Bush--Mosteller learning~\cite{Bush1955book}, and social learning~\cite{fudenberg1998book,mcelreath2008book}---approach the paradigmatic replicator equation~\cite{Taylor1978mathbio} under certain limits~\cite{,Borgers1997JET,Hopkins2005GEB,Beggs2005JET}. 

The replicator equation is one of the cornerstones of the deterministic frequency-dependent models of microevolution studied as part of evolutionary game theory~\cite{SMITH1973nature,smith1993theory}. Interestingly, the replicator equation transcends the boundary between the biological evolution and reinforcement learning process to find application in both the fields. In fact, the time-discrete form of the replicator equation can directly be interpreted as a Bayes' update rule~\cite{harper2009arxiv}. Furthermore, mutations in evolution may be seen analogous to mistakes made in a learning process.

Making a few mistakes during learning is practically inevitable and it usually has a negative connotation attached to it. But in many real-life situations, we can find mistakes facilitating new opportunity to handle a situation more comfortably. This is analogous to the case of biological evolution where mutations---mistakes in DNA replication---help organisms to evolve and adapt to  ever changing environment. Experimental studies showed that foraging errors play a positive role in resource exploration
by bumblebees~\cite{Evans2014JCPA}. In the context of education and industry, learning through mistakes takes an significant role~\cite{Tulis2016FLR,Cyr2015JEPLMC,vanlehn1988book}: Recent studies showed that to get the optimal performance in high-stakes situations students should be encouraged to commit and correct errors when they are in low-risk learning situations~\cite{Metcalfe2017ARP}. Furthermore, the Japanese active learning approach~\cite{stevenson1994book} where students first try to solve problems on their own---a process that is likely to be filled with false starts---is based on the idea of learning-from-errors. It is capable of rendering better performance as compared to the teaching methodology based on avoidance of errors.

Mistakes can be easily modelled into the replicator equation which modifies to replicator-mutator equation~\cite{Nowak2001science,KOMAROVA2001jtb,Mobilia2010jtb,Pais2011ieee,Toupo2015PRE,You2017pre,Mittal2020pre,Mukhopadhyay2021JoPC}. At the very least, there are two ways in which the change in the probability of playing a strategy in a subsequent round can be effected: additive and multiplicative mutations. While the multiplicative mutation is more biological in nature---the error in DNA replication directly effected the offsprings' type~\cite{nowak2006evolutionary}, the additive mutation has been advocated to be more social in nature---it occurs directly in the adults~\cite{Mobilia2010jtb, Toupo2015PRE}. However, the former has been used in social settings like in the evolution of language~\cite{Nowak2001science}, whereas the latter has been used in biological settings~\cite{rice2004evolutionary}.

An important feature of the replicator equation (also, the replicator mutator equation) is that it not only leads to convergent outcomes, but also predicts non-convergent oscillatory (periodic and chaotic) solutions. The omnipresent game of Rock-Paper-Scissors (RPS)~\cite{walker2004ss,livingstone2008cup} does exhibit oscillatory outcomes by virtue of non-transitive dominance relation between the pure strategies. Omnipresence of RPS is amazing: It is found in humans~\cite{Semmann2003nature}, birds~\cite{Jukema2006bl}, reptiles~\cite{Sinervo1996nature}, insects~\cite{Chippindale2013me}, marine beings~\cite{Shuster1991nature}, microbes~\cite{Kirkup2004nature}, and many other scenarios. How learning is effected by quenching the cyclic outcomes is a curious question in this setup. 

Putting our aim in a little bit more technical terms, reinforcement learning individuals have access to different strategies and corresponding to each strategy they get a specific payoff. Under the assumption that the players are rational, we expect that repeated playing should lead them to the Nash-equilibrium (NE)~\cite{nash1951am,Kalai1993eco,holt2004pnas}. In this paper, we shall say that a player has learnt if they end up playing NE after repeated trials~\cite{Ianni2014JME}. Since NE corresponds to a fixed point of replicator dynamics~{\cite{Cressman2014pnas}}, the asymptotic stability of the fixed point is synonymous to learning for us. RPS and its generalizations are interesting because not only periodic orbits~\cite{Schuster1981bc} but also chaotic trajectories~\cite{Sato2002pnas} are seen under the replicator dynamics. Taming these non-convergent solution via mistake or mutation is a clear possibility. In this evolutionary game-theoretic context, it is worth mentioning that mistakes were shown to help establish signalling system {\cite{Hofbauer2008JTB}}, to quench oscillatory outcomes~{\cite{Kleshnina2021plos}}, and to allow the appearance of evolutionarily stable strategies in repeated prisoners' dilemma game~\cite{Boyd1989jtb}.

Without further ado, we now introduce the replicator-mutator equations (as obtained, in the Appendices, from some standard learning mechanisms) and present their rather extensive investigation for RPS game. The main focus is on how mistakes can facilitate learning. 

\section{Dynamical Equations} 
Replicator equation is used in various settings---Darwinian evolution and reinforcement learning being the ones we are explicitly interested herein. For example, consider the scenario of two players repeatedly playing a one-shot game where one player chooses the $j$th strategy (out of $J$ number of pure ones available to her) with frequency $x_j$ and the second player, who has access to $K$ pure strategies, plays $k$th strategy with frequency $y_k$. The payoffs of the game interaction between the two players are summarised in payoff matrices ${\sf A}$ and ${\sf B}$, respectively for the first and the second players. An advantageous pure strategy is reinforced in the subsequent steps through a learning mechanism that in the continuous time limit, many a times,  boils down to~\cite{Sato2002pnas,hopkins2002econometrica,Beggs2005JET,Borgers1997JET,borgers2000IER} the replicator-mutator equation:
\begin{subequations}\label{17}
	\begin{eqnarray}
	&&\frac{dx_j}{dt}=x_j[(\mathbf{{\sf A}y})_j-\mathbf{x}^T\mathbf{{\sf A}y}]+\sum_{\substack{i=1\\i\ne j}}^{J}\mu^x_{ji}x_i-\sum_{\substack{i=1\\i\ne j}}^{J}\mu^x_{ij}x_j,~~~~~~~~~~ \\&&
	\frac{dy_k}{dt}=y_k[(\mathbf{{\sf B}x})_k-\mathbf{y}^T\mathbf{{\sf B}x}]+\sum_{\substack{i=1\\i\ne k}}^{K}\mu^y_{ki}y_i-\sum_{\substack{i=1\\i\ne k}}^{K}\mu^y_{ik}y_k. 
	\end{eqnarray}
\end{subequations}
Here we have allowed for mistakes as well~(see Appendix~\ref{a1}):  the non-negative $\mu^x_{ij}$ ($\mu^y_{ij}$) is the rate with which the first (second) player mistakenly plays $i$th pure strategy when she should have played the $j$th one. To clarify the symbols used, we mention that $\mathbf{x}$ corresponds to column vector for frequency $x_i$, the first term inside the square bracket $\mathbf{({\sf A}y})_i$ corresponds to the expected payoff of the $i$th type of individuals and the second term $\mathbf{x}^T\mathbf{{\sf A}y}$ corresponds to the average payoff of the player one. The symbols in the equation for the second player may be interpreted likewise. 

Interestingly, there is another way one could have incorporated mistakes in the replicator-mutator equation, viz.,
\begin{subequations}\label{18}
	\begin{eqnarray}
	&& \frac{d x_j}{dt}=\sum_{i=1}^{J}x_i(\mathbf{{\sf A}y})_i q^x_{ji}-x_j({\bf x}^T \mathbf{{\sf A}y}),\\&&
	\frac{dy_k}{dt}=\sum_{i=1}^{K} y_i (\mathbf{{\sf B}x})_i q^y_{ki}-y_k ({\bf y}^T \mathbf{{\sf B}x}),
	\end{eqnarray}
\end{subequations}
where $q^x_{ij}$ ($q^y_{ij}$)---elements of a row stochastic matrix---corresponds to the probability of making mistake  from strategy $j$ to strategy $i$ by the player one (two). The form of mutation considered here is called multiplicative mutation in contrast to `additive mutation'---a term used for the earlier case.

The equations above can be interpreted in the setting of biological microevolution as well: Consider a \emph{non-homogeneous} population, such that it consists of two different classes of individuals; each class of individuals has a particular role. Class one has individuals of $J$  different (pheno-)types, while an individual of the other class can be of one of the $K$ possible types. We denote the frequencies of the types in class one by $x_1, x_2,\cdots,x_J$, and similarly, the frequencies of the types in class two are denoted by $y_1, y_2,\cdots,y_K$. Every individual of one class can randomly interact with every other individuals of the other class to obtain a payoff (measuring the reproductive success) that is represented as an element of a $J \times K$ payoff-matrix, $\textsf{A}$; similarly, the payoff matrix for the second class is a $K \times J$ matrix, $\textsf{B}$. Note there is no intraclass interaction considered, or even if there is any, it is supposed to not yield any payoff. If the population is, furthermore, infinite, well-mixed, generation-wise overlapping, and haploid (biologically speaking), then the aforementioned replicator-mutator equation is a paradigmatic model of the replication-selection process with the rates/probabilities of mistakes should now be identified as mutations. 

In fact, when the population is homogeneous (single class with intraclass interations leading to payoff matrix ${\sf A}$), if there are $J$ possible types of individuals and their frequencies be $x_1, x_2,..., x_J$ respectively, then we arrive at the more well-known form of equation called the replicator equation~{\cite{Taylor1978mathbio}}:
\begin{equation}\label{3}
\dot{x}_i=x_i[(\mathbf{{\sf A}x})_i-\mathbf{x}^T\mathbf{{\sf A}x}],~~~~~~~~~\forall i \in \{1,2, \cdots,J\},
\end{equation}
 in the absence of any mutation. In the presence of additive or multiplicative mutations, this gets respectively modified to
\begin{subequations}
\begin{eqnarray}
\label{6}
\dot{x}_i&=&x_i[(\mathbf{{\sf A}x})_i-{\bf x}^T\mathbf{{\sf A}x}]+\sum_{\substack{j=1\\j\ne i}}^{J}\mu_{ij}x_j-\sum_{\substack{j=1\\j\ne i}}^{J}\mu_{ji}x_i,\quad\\
\label{7}
\dot{x}_i&=&\sum_{j=1}^{J}q_{ij}x_j(\mathbf{{\sf A}x})_j-x_i({\bf x}^T\mathbf{{\sf A}x}),\label{eq:qrme}
\end{eqnarray}
\end{subequations}
$\forall i\in\{1,2,\cdots J\}.$ Here, the non-negative $\mu_{ij}$ is the rate of mutation from $j$ to $i$ strategy, while $q_{ij}$---$ij$-th element of a row stochastic matrix, $\sf Q$, say---is the probability of mutation from $j$ to $i$ strategy. Biologically speaking, the multiplicative is apposite if the mutations take place in the offsprings but if they do in the adults then the additive one makes sense---hence, the latter may be called social mutation as it may happen due to behavioural changes post social interactions. However, the additive mutation is used to model the mutations in offspings as well~\cite{rice2004evolutionary}.

The replicator-mutator equation for \emph{homogeneous} well-mixed population can also be interpreted differently in the context of learning, specifically, the social learning~({\cite{mcelreath2008book,fudenberg1998book}}; see Appendix~\ref{app-3}). Social learning is a kind of learning in which individuals of a society update their behaviours by adopting others' seemingly beneficial behaviour. Social learning can be at the level of idea, knowledge, philosophy, religion, political views, and many other social or cultural contexts. Not only in humans but in the animal world also, social learning is a common phenomenon~\cite{jablonka2014mit}. Furthermore, Eq.~(\ref{7}) has also been used in the context of evolution of universal grammar {\cite{Nowak2001science}}.

\section{The Questions}
\begin{figure*}
	\centering
	\includegraphics[scale=0.6]{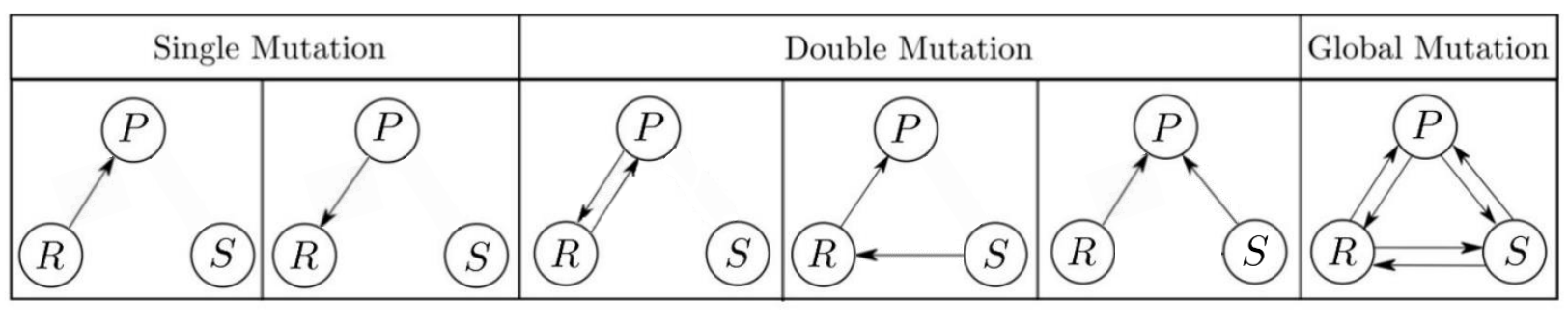}
	\caption{Different motifs of mutation pattern in RPS game considered in this paper. Arrows indicate the direction of mutations, i.e., $R\xrightarrow[]{}P$ means that $R$ can mutate to $P$.}
	\label{fig:my_label11}
\end{figure*}
Toupo and Strogatz~{\cite{Toupo2015PRE}} studied replicator dynamics [Eq.~(\ref{3})] in the presence of additive mutation when the individuals of the population are involved in Rock-Paper-Scissors (RPS) game~\cite{walker2004ss} whose payoff matrix was chosen in the following form:
\begin{equation}
{\sf A} \equiv
\begin{bmatrix}
0 & -(1+\epsilon) & 1\\
1 & 0 & -(1+\epsilon)\\
-(1+\epsilon) & 1 & 0
\end{bmatrix},  
\label{eq:rps1}
\end{equation}
where $\epsilon$ can lie in the range $(-1,\infty)$. The choice of this range effectively excludes the apostatic RPS games~\cite{friedman2016oup}. Obviously, the only NE equilibrium corresponding to this payoff matrix is $(1/3,1/3,1/3)$. Some quite interesting dynamical outcomes were reported. Many different motifs of mutations were considered---some illustrated in Fig.~\ref{fig:my_label11}---and, among others, following results were found.

In the presence of global mutation only one fixed point exists: Internal fixed point, $(1/3,1/3,1/3)$, is stable only for higher mutation values ($\mu>\epsilon/18$), while for  $\mu<\epsilon/18$ limit cycle appear via the Hopf bifurcation at $\mu=\epsilon/18$ rendering the realization of the NE point---in other words, the learning---impossible at the lower mistake-rates. In the case of single mutation where the mutation happens unidirectionally from one strategy to another that can beat the former, e.g., $R$ to $P$, a total of three fixed points appear: two corner ones and one internal which is close to NE for small mutation. Here also, at the lower mutation rates, convergence to the internal fixed point is disrupted due to appearance of Hop-bifurcation-mediated limit cycles. When the mutation is in the opposite direction, i.e., $P$ to $R$, an additional fourth fixed point appears on the boundary. In this case also, at lower values of mutation, the inner fixed point loses its stability to a limit cycle through the Hopf bifurcation. For higher values of mutation, however, a coexistence region of $R$ and $P$ can appear due to the transcritical bifurcation of the interior fixed point: The fixed point stably exists only at intermediate rates of mistake. Exactly same situation is observed in the opposing mutation---where a particular strategy mutates out of the other two at the same rate: e.g., $R$ to $P$ and simultaneously, $S$ to $P$. Whereas, in the bidirectional mutation---a form of double mutations where there is symmetric bidirectional mutation between any of the two strategies, say, $R$ and $P$---situation about the inner fixed point is identical to the case of global mutation except that the Hopf bifurcation occurs at $\mu=\epsilon/6$. Lastly, in the double mutation in the direction of circulation (see Fig.~\ref{fig:my_label11}; e.g., $S$ to $R$ and simultaneously, $R$ to $P$), the inner fixed point (which is close to NE for small mutation) looses its stability to a Hopf-bifurcation-mediated limit cycle. But for higher values of mutation, the internal fixed point becomes locally asymptotically stable.

In an unrelated work by Sato et al.~\cite{Sato2002pnas}, reinforcement two-player learning dynamics under zero-sum RPS game was studied. Specifically, they took the payoff matrices  
\begin{equation}
{\sf A}=\begin{bmatrix}
\epsilon_x & -1 & 1\\
1 & \epsilon_x & -1\\
-1 & 1 & \epsilon_x
\end{bmatrix}~{\rm and}~
{\sf B}=\begin{bmatrix}
\epsilon_y & -1 & 1\\
1 & \epsilon_y & -1\\
-1 & 1 & \epsilon_y
\end{bmatrix}, \label{eq:AB}
\end{equation}
where $-1<\epsilon_x<1$, $-1<\epsilon_y<1$, and when the game is specifically considered zero-sum game, one also has $\epsilon_x=-\epsilon_y=\epsilon$. This bimatrix game has a single NE equilibrium which corresponds to ${\bf x}=(1/3, 1/3, 1/3)$ and ${\bf y} = (1/3, 1/3, 1/3)$. They showed that the system becomes a Hamiltonian system---even Hamiltonian chaos can appear---so that no asymptotically stable fixed point is possible. Thus, the dynamics never can reach the NE point, and learning is not possible at all. 

We note that while the systematic effect of mutation/mistake was not at all done in the work by Sato et al., the effect of the multiplicative mutation remained unaddressed in the work by Toupo and Strogatz. This is exactly where our present work comes in. Before we spell out the specific questions, one subtlety that makes our work very interesting must be mentioned. When dealing with the additive mutation, any constant shift in the payoff matrices (in the case of aforementioned bimatrix games, it leads to constant sum games), keeps the replicator-mutator equation invariant. But this is not so when multiplicative mutation is invoked. In fact, if all the elements of the payoff matrices are not chosen to be positive, the dynamics may not even be confined within the simplex: the frequencies can becomes either negative or greater than unity as time flows---rather unphysical solutions~(see Appendix~\ref{app:fi} for details). We thus can choose a shift parameter to make the payoff matrices positive so that we can contrast the effects of the multiplicative mutation with the additive mutation on the replicator-mutator dynamics. However, we need to be careful because shift of payoff matrix may completely change the dynamical outcomes under multiplicative mutation. So while performing the comparison we need to consider dynamical outcomes for different values of the shift parameter. Mathematically, another way by which we can make the system forward invariant for any payoff matrix is to impose a boundary condition which forcefully restricts the system inside the physical region: whichever frequency becomes unphysical is stopped from evolving any further (detailed in Sec.~\ref{section_6}). We call this `impenetrable boundary condition'. But this boundary condition can also effect the asymptotic states of the system.

The main questions, we ask in this {paper} are as follows:
\begin{itemize}
	\item What is the effect of multiplicative mutation on the {$RPS$}-game being played in a homogeneous population? How does it compare against additive mutation? What is the effect of the shift parameter needed to make the system under multiplicative mutation forward invariant?
	\item How the results, thus, found change as the impenetrable boundary condition is imposed?
	\item How does the scenario of bifurcations compare in the cases of multiplicative and additive mutations?
	\item Do mutations/mistakes preserve the Hamiltonian chaos?
	\item If and when the mutations/mistakes disrupt/tame the Hamiltonian chaos, do they lead to dissipative chaos or periodic oscillatory solutions?
	\item What are the physical interpretations of the dynamical outcomes arrived at?
	\end{itemize}
In what follows, for the sake of lucid and uncluttered presentation without losing essence of our claims, let us consider only global mutation in detail and for other cases, we shall succinctly summarize the results.

\section{Homogeneous population} 
In the case of homogeneous population with no separate classes of individuals in different roles, as we mentioned in the previous section, the results for the replicator-mutator dynamics with additive mutation are already known. Here we focus on the effect of the multiplicative mutation. We choose the mutation matrix corresponding to the global mutation case:
\begin{equation}\label{16}
\sf{Q}=
\begin{bmatrix}
1-q & q/2 & q/2\\
q/2 & 1-q & q/2\\
q/2 & q/2 & 1-q
\end{bmatrix}
\end{equation}
which is both column and row stochastic. 

\begin{figure}
	\centering
	\includegraphics[scale=0.4]{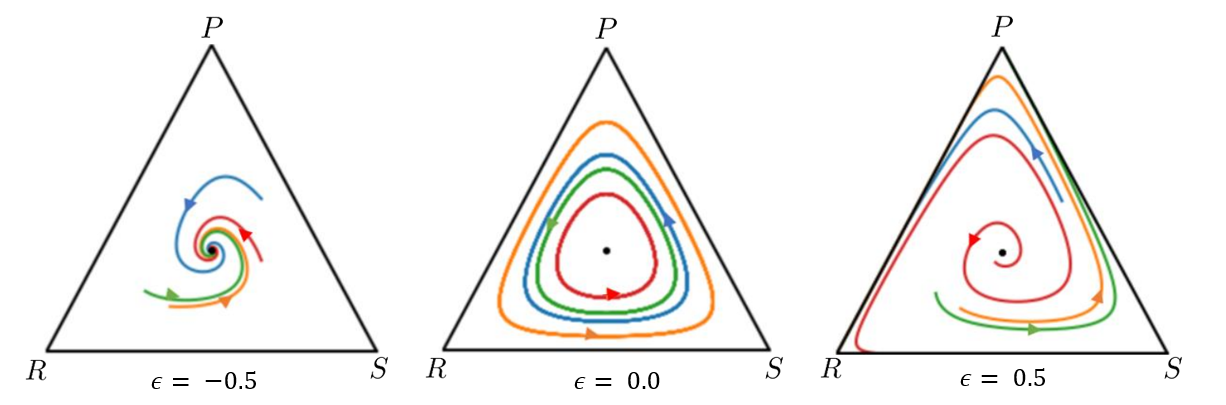}
	\caption{Degenerate Hopf bifurcation renders the internal fixed point unstable. The dynamics spiral outward and eventually stabilize at the boundary, due to the presence of impenetrable boundary condition: We consider multiplicative mutation with impenetrable boundary and fix $q=0.2$. For $\epsilon<0,$ (leftmost subplot) the inner fixed point is asymptotically stable. At $\epsilon=0$ (central subplot), the degenerate Hopf bifurcation occurs and the fixed point becomes non-hyperbolic. Finally, orbits spiral out from the vicinity of internal fixed point and asymptotically approach the boundary for $\epsilon>0$ (rightmost subplot).}
	\label{fig:my_label1}
\end{figure}
On solving the replicator-mutator equation [Eq.~(\ref{eq:qrme})] using the above mutation matrix and payoff matrix $A$ [see Eq.~(\ref{eq:rps1})], we get four possible fixed points: $(0,1,0)$, $(0,0,1)$, $(1,0,0)$, and $(1/3, 1/3, 1/3)$. The linear stability analysis around the fixed point $(1/3,1/3,1/3)$, which corresponds to NE for the payoff matrix, gives following eigenvalues:
\begin{equation}
\lambda_{\pm}=\frac{1}{12}\left[(2 + 3q) \epsilon \pm i\sqrt{3}(2 + \epsilon) (3q - 2)\right].
\end{equation}
A degenerate Hopf bifurcation occurs at $\epsilon=0$ (genericity condition is violated, i.e., the first Lyapunov coefficient is identically zero~\cite{Kuznetsov2006Scholarpedia}): a family of nonisolated closed orbits appear about the fixed point. For $\epsilon < 0$, the fixed point is a spiral node; and for $\epsilon > 0$, trajectories end up onto the boundary of simplex (see FIG.~\ref{fig:my_label1}). This behaviour is same for all values of $q$ except $q=2/3$ where the eigenvalues become real, but the stability condition remains the same, i.e., for $\epsilon<0$ the fixed point is stable, and for $\epsilon>0$ it is unstable. So overall, we can say the perfect learning of reaching the NE is possible for all values of mutation when $\epsilon<0$. the perfect learning of reaching the NE is possible for all values of mutation when $\epsilon<0$. This is unlike what happens in the additive case where the threshold epsilon is mutation rate dependent. Note that since in the case under consideration some payoff elements are negative, we must impose impenetrable boundary condition for getting the global phase portrait.

\begin{figure}
	\centering
	\includegraphics[scale=0.5]{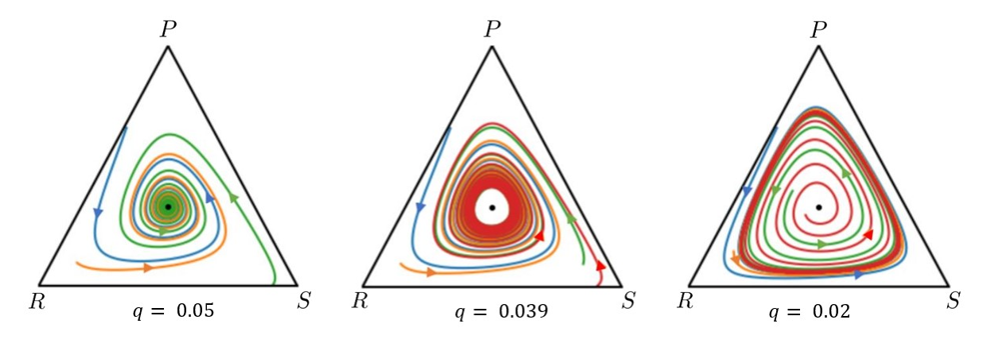}
		\caption{{Hopf bifurcation induced limit cycle: We consider multiplicative mutation with payoffs shifted by $c=3$. We keep $\epsilon=1$. For $q>0.039$ (leftmost subplot) the inner fixed point is asymptotically stable. At $q\approx0.039$ (central subplot), the Hopf bifurcation occurs (notice the slowing down of the approach of the orbit onto the fixed point) making the fixed point non-hyperbolic. Finally, the limit cycle is seen for $q<0.039$ (rightmost subplot).}}
	\label{fig:my_label3}
\end{figure}
\begin{figure}
	\centering	
	\includegraphics[scale=0.2]{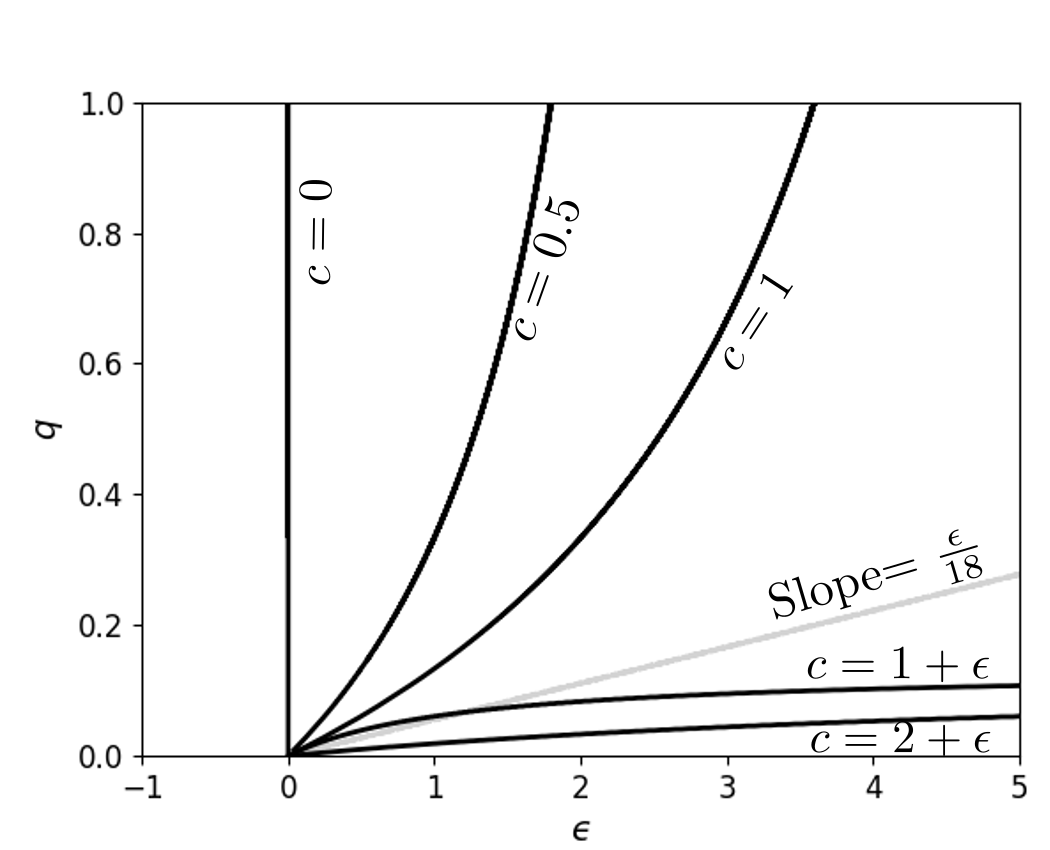}
	\caption{The demarcation of stable and unstable regions in $\epsilon$-$q$ parameter space for the internal fixed point.
 The regions below respective black lines correspond to limit cycle about the unstable internal fixed point, while the region above (in the first quadrant and the second quadrant) corresponds to stable internal fixed point. For comparison, we have plotted a gray line that acts as the boundary between stable and unstable regions in the case of additive mutation.}
	\label{fig:boat4}
\end{figure}

If we want to avoid imposing the boundary condition, then we have to choose a shifted payoff matrix with a shift parameter $c$:
\begin{equation}
{\sf A}_c \equiv
\begin{bmatrix}
c & c-\epsilon-1 & 1+c\\
1+c & c & c-\epsilon-1\\
c-\epsilon-1 & 1+c & c
\end{bmatrix},
\end{equation}
where the shift parameter satisfies the relation---$c \ge (1+\epsilon)$. Dynamical equation (\ref{7})  under RPS game and shift parameter takes the following form:
\begin{subequations}\label{23}
	\begin{eqnarray}
	&&\frac{d x_1}{dt}=\sum_{i=1}^{3}x_i(\mathbf{{\sf A}x})_i q_{1i}-x_1(\mathbf{x}^T \mathbf{{\sf A}x})\nonumber\\
	&&\phantom{\frac{d x_1}{dt}=}+c\left[\sum_{\substack{i=2}}^3 q_{1i} x_i-\sum_{\substack{i=2}}^3 q_{i1} x_1\right],\\
	&& 
	\frac{dx_2}{dt}=\sum_{i=1}^{3}x_i(\mathbf{{\sf A}x})_i q_{2i}-x_2(\mathbf{x}^T \mathbf{{\sf A}x})\nonumber\\
	&&\phantom{\frac{d x_1}{dt}=}+c\left[\sum_{\substack{i=1\\i\ne 2}}^3 q_{2i} x_i-\sum_{\substack{i=1\\i\ne 2}}^3 q_{i2} x_2\right].
	\end{eqnarray}
\end{subequations}
Interestingly, the form of the equation (\ref{23}) suggests that the effect of shift parameter is effectively as if both the additive and multiplicative mutations are present simultaneously, {where the effective additive mutation should be recognized as $cq_{ij}$}. Again the linear stability analysis about the NE---which is the {only fixed point $(1/3,1/3,1/3)$}---yields eigenvalues for the Jacobian at the fixed point as:
\begin{equation}
\lambda_{\pm}=\frac{1}{12}\left[\epsilon(2 + 3q)-18cq \pm i\sqrt{3}(2 + \epsilon) (3q - 2)\right].
\end{equation}
For $c> (3q+2)\epsilon/18q$, the fixed point is globally stable: learning always happens for any initial strategy. For $c< (3q+2)\epsilon/18q$, the fixed point becomes unstable and a stable limit cycle appear via the Hopf bifurcation at $c= (3q+2)\epsilon/18q$ (see FIG.~\ref{fig:my_label3}), rendering the learning impossible. The result is not that qualitatively different from the additive case except that the learning depends explicitly on the shift parameter only in the multiplicative mutation case. Of course, now the threshold beyond which attainment of the NE is possible depends explicitly on the value of $c$. We exhibit this in Fig.~\ref{fig:boat4} which summarily depicts that the higher value of $c$, the lower is threshold of mistakes for reaching the NE.

\section{Non-homogeneous Population}
\begin{figure*}
	\centering
	\includegraphics[scale=0.65]{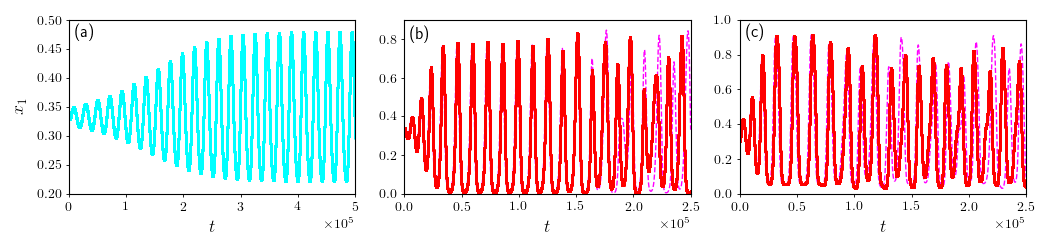}
	\caption{Non-convergent outcomes in non-homogeneous population with symmetric payoff matrices. We observe oscillatory outcome in global additive mutation ($\mu=0.05$ and $\epsilon=-0.97$; subplot (a)), chaotic outcome in additive single mutation from {Rock to Paper} ($\mu=0.05$ and $\epsilon=0.75$; subplot (b)) and chaotic outcome in multiplicative single mutation from {Paper to Rock} ($q=0.05$, $\epsilon=0.75$ and $c=1.75$; subplot (c)). The red time-series and the magenta time-series evolve from two neighboring initial conditions.}
	\label{fig:b1}
\end{figure*}

What if the population is non-homogeneous in the sense that there are two classes with distinct roles with only interclass competition? The easiest extension is that of RPS game itself but now we use Eqs.~(\ref{17}) and (\ref{18}) with ${\sf A}={\sf B}^T$. As showcased in Fig.~(\ref{fig:b1}), with different modes of mutations/mistakes periodic and chaotic solutions appear, implying that the learning does not converge.

As summarized earlier in this paper, chaotic dynamics---specifically Hamiltonian chaotic dynamics---in the context of reinforced learning through the replicator equation of Bimatrix zero-sum {RPS} games [with $\sf A$ and $\sf B$ given in Eq.~(\ref{eq:AB})] is well-documented~{\cite{Sato2002pnas}}; with $\epsilon_x\ne\epsilon_y$ one gets other non-convergent dynamical structures, like heteroclinic cycle. In this section, we confine ourselves to the global mutation case only: $\mu^x_{ij}=\mu_x$ and $\mu^y_{ij}=\mu_y$ $\forall i,j$ for additive case, while for the multiplicative case both $q_{ij}^x$ and $q_{ij}^y$ are the appropriate matrix elements of $\sf Q$ [see Eq.~(\ref{16})]. 

\subsection{\textbf{Additive Mutation}}
Dynamical equation (\ref{17}) has only a single fixed point--- $({\mathbf{x}^*},{\mathbf{y}^*})=(1/3, 1/3, 1/3,1/3, 1/3, 1/3)$---which corresponds to NE. The linear stability analysis about the fixed point $({x*}, {y*})$ yields the two degenerate eigenvalues:
		\begin{eqnarray}
		&& \lambda_{\pm}=\frac{1}{6}\left[-9(\mu_x+\mu_y)\pm i\sqrt{12+4 \epsilon^2-81(\mu_x-\mu_y)^2}\right]\nonumber.\\\label{12}
		\end{eqnarray}
The fixed point can become unstable only if the expression inside  square-root is negative and greater than the square of the term outside square-root, i.e.,
\begin{subequations}
	\begin{eqnarray}
	&& 81(\mu_x-\mu_y)^2-12-4 \epsilon^2>81 (\mu_x+\mu_y)^2\qquad\\
	\implies &&\mu_x\mu_y<-\frac{(3+\epsilon^2)}{81},
	\end{eqnarray}
\end{subequations}
which is never allowed as the mistake rates are non-nagative.
This implies that fixed point $({\mathbf{x}^*}, {\mathbf{y}^*})$ is always globally stable, i.e., perfect learning always enforced in presence of global additive mistake. Note that even if one player is \emph{perfect}---defined to be the one who makes no mistakes---i.e., either $\mu_x$ or $\mu_y$ is zero, the same conclusion holds.
%
	
\subsection{\textbf{Multiplicative Mutation}}
So we note that the additive mutation destroys the Hamiltonian structure completely and chaos, if any, is suppressed to lead to learning. However, the case of multiplicative mutation is richer as we shall see soon.

\subsubsection{Shifted Payoff Matrix}
First we rewrite dynamical equation (\ref{18}) in the presence of shifted payoff matrices---to ensure forward invariability---obtained by adding a parameter $c$ to every element of $\sf A$ and $\sf B$:
	\begin{subequations}\label{28}
		\begin{eqnarray}
		&& \frac{d x_j}{dt}=\sum_{i=1}^{3}x_i({\sf A}{\bf y})_i q^x_{ji}-x_j({\bf x}^T {\sf A}{\bf y})\nonumber\\ &&\phantom{\frac{d x_j}{dt}=}+c\left[\sum_{\substack{i=1\\i\ne j}}^3 q_{ji}^x x_i-\sum_{\substack{i=1\\i\ne j}}^3 q_{ij}^x x_j\right],\\
&&\frac{dy_k}{dt}=\sum_{i=1}^{3} y_i ({\sf B}{\bf x})_i q^y_{ki}-y_k ({\bf y}^T {\sf B} {\bf x})\nonumber\\
&&\phantom{\frac{d x_j}{dt}=}+c\left[\sum_{\substack{i=1\\i\ne k}}^3 q_{ki}^y y_i-\sum_{\substack{i=1\\i\ne k}}^3 q_{ik}^y y_k\right],
		\end{eqnarray}
	\end{subequations}
$\forall j,k\in\{1,2\}$. 

\begin{figure}
	\centering	
	\includegraphics[width=70mm, height=50mm]{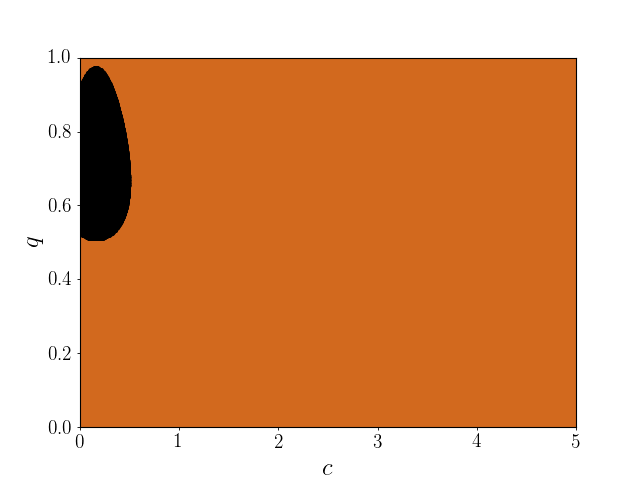}
	\caption{Learning in two player bimatrix game with multiplicative mutation. The figure exhibits $c$-$q$ parameter space's regions that correspond to either asymptotically stable NE (orange region) or unstable NE (black colour). Here, $\epsilon=0.5$.}
	\label{fig:boat3}
\end{figure}

When the mutation matrix $\sf Q$ [see Eq.~(\ref{16})] is chosen for both the players, we see that NE corresponds to the interior fixed point, linear stability about which gives the folllowing eigenvalues,
\begin{subequations}
	\begin{eqnarray}
	&&\lambda_{1}^\pm=\frac{1}{3}\left(-3cq\pm\sqrt{\alpha_1-6\sqrt{\beta_1}}\right),\\
	&&\lambda_{2}^{\pm}=\frac{1}{3}\left(-3cq\pm\sqrt{\alpha_1+6\sqrt{\beta_1}}\right),
	\end{eqnarray}
\end{subequations}
where $\alpha_1\equiv-3-\epsilon^2+9q+3q\epsilon^2-27q^2/4+9c^2q^2/4$, $\beta_1\equiv-c^2\epsilon^2q^2/4+3c^2\epsilon^2q^3/4$. One can directly check that for small values of $q$ the fixed point is always stable; for other values of $q$ how the stability changes with shift parameter $c$ can be concluded from FIG.~\ref{fig:boat3}. We observe that the range of $q$-values for which the NE is asymptotically stable increases with the increase in the value of the shift parameter. If we realize that the form of Eq.~(\ref{28}) mimics the simultaneous presence of both the additive and multiplicative mutations, where the effective additive mutation rate is given by $cq_{ij}$, we can as well conclude that the simultaneous presence of multiplicative and additive mistakes can facilitate learning.  

\begin{figure}
	\centering	
	\includegraphics[width=90mm, height=40mm]{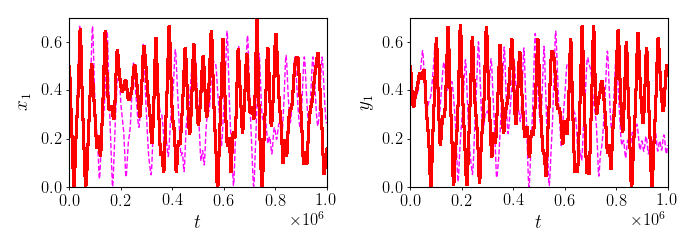}
	\caption{Chaos due to multiplicative mutation with impenetrable boundary condition. The left and the right subplots show the chaotic outcomes for the two players' strategies. The red time-series and the magenta time-series evolve from two neighboring initial conditions. Here, $q=0.84$ and $\epsilon=0.5$.}
	\label{fig:boat2}
\end{figure}
\begin{figure}
	\centering	
	\includegraphics[width=90mm, height=70mm]{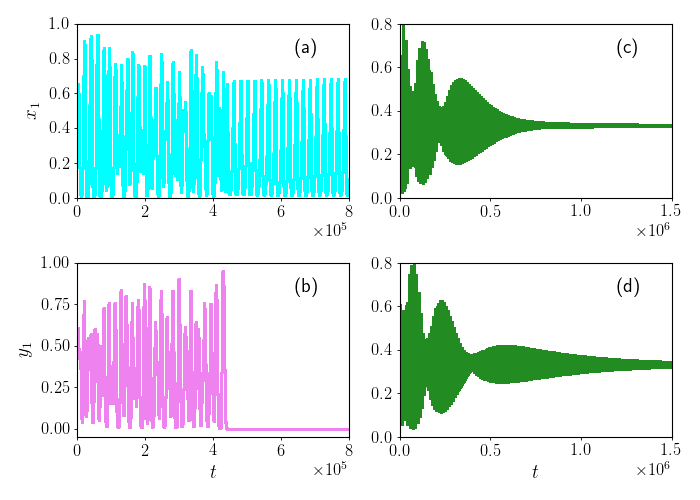}
	\caption{One perfect player makes the learning possible. We take $q=0.01$ and $\epsilon=0.5$ in the mutiplicative mutation case with impenetrable boundary. When none of the two players are perfect (subplots (a) and (b)), the dynamics does not reach the NE. When only player one is not perfect (subplot (c)), the other player (subplot (d))---along with the player one---reaches the NE asymptotically.}
	\label{fig:boat1}
\end{figure}
\begin{figure}
	\centering	
	\includegraphics[scale=0.2]{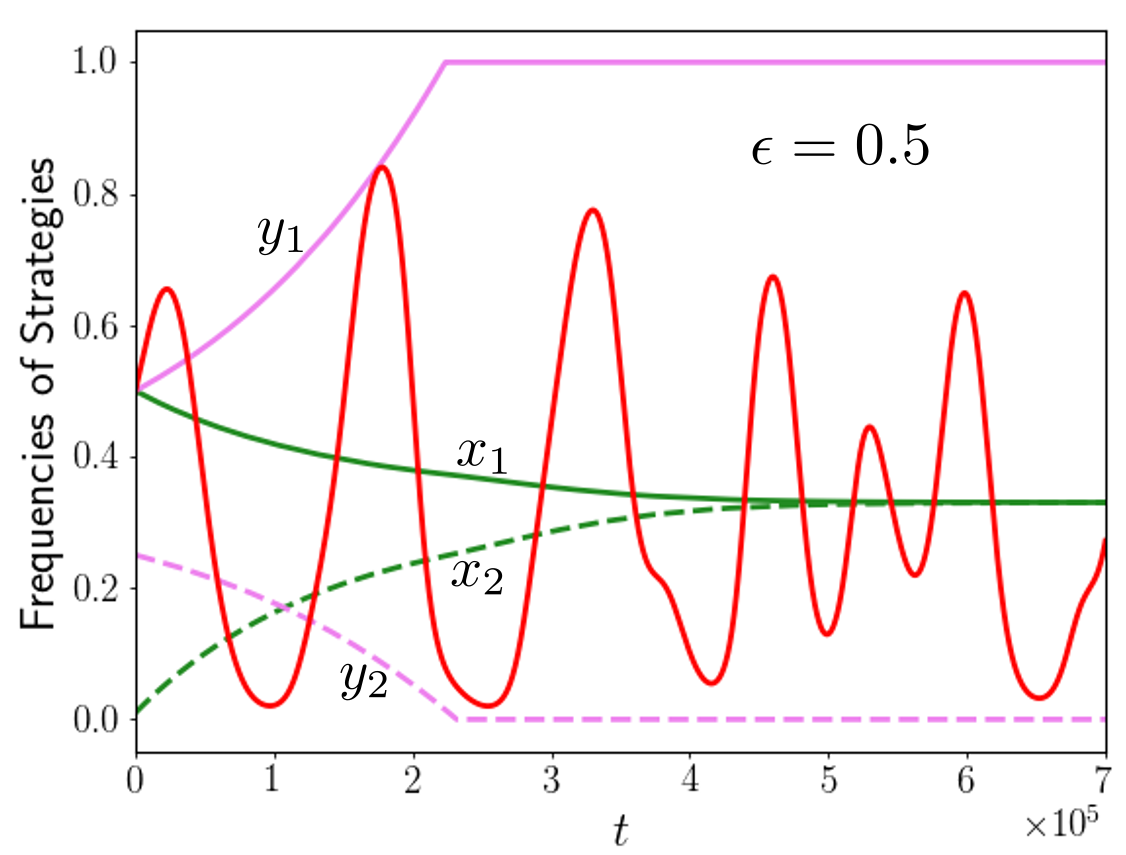}
	\caption{Mistakes tame chaos in the Hamiltonian zero-sum bimatrix game. The red curve is the chaotic trajectory in the absence of any mistake ($q=0$). With $q=0.33$ and imposing impenetrable boundary condition, the frequencies of strategies of the two players---green and magenta curves respectively for player one and player two---reach the NE asymptotically. We have 
added unity to the elements of $\sf A$ and subtracted the same from the elements of $\sf B$.}
	\label{fig:boat5}
\end{figure}   
\begin{figure*}
	\centering	
	\includegraphics[scale=0.3]{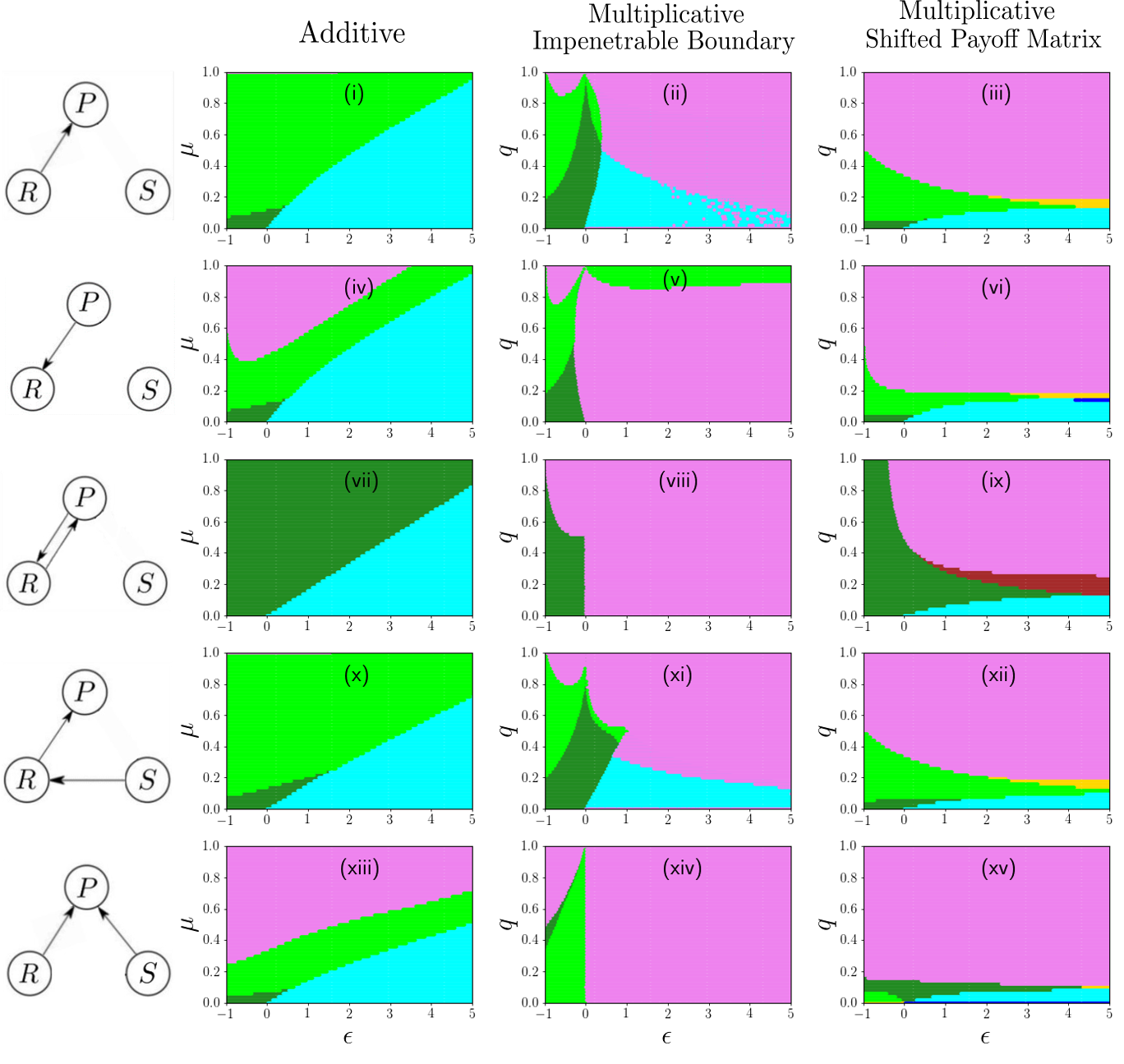}
	\caption{Classification of asymptotic outcomes in $\epsilon$-$\mu$ (and $\epsilon$-$q$) parameter space  under different single and double mutation motifs in homogeneous population. The forest-green, the lime, the cyan, and the magenta regions respectively denote stable internal fixed point within the circular neighbourhood of radius 0.1 about the NE, stable internal fixed point but outside the circular neighbourhood of radius 0.1 about the NE, oscillatory limit cycle solution, and stable boundary point. The brown and the gold regions represent the simultaneous existence of stable internal (brown color for NE and gold colour for  non-NE) fixed point and boundary stable fixed point. In the last column of the figure, the shift parameter is fixed as $c=1+\epsilon$.}
	\label{fig:table}
\end{figure*}
\begin{figure}
	\centering
	\includegraphics[scale=0.18]{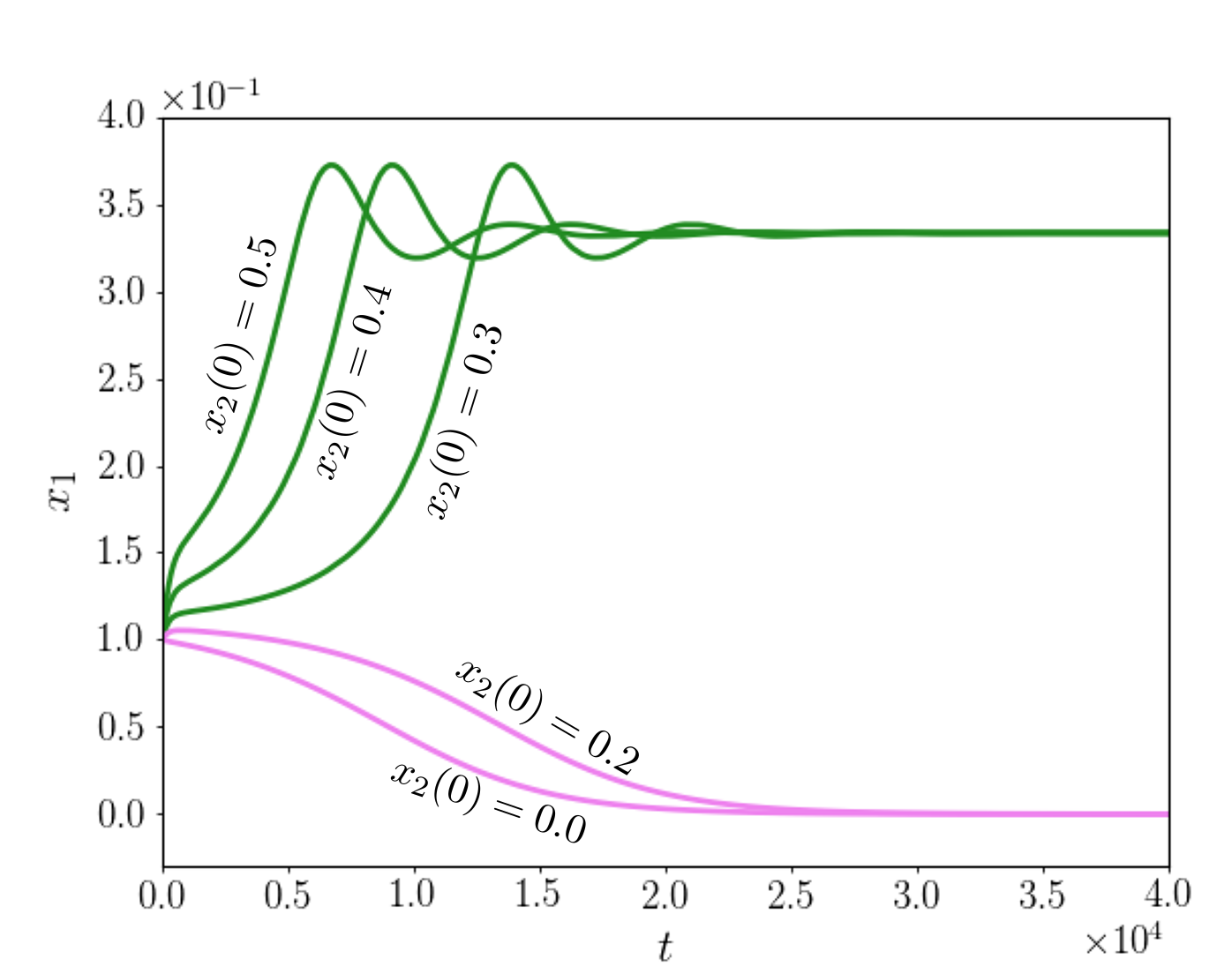}
	\caption{Coexistence of the internal and the boundary fixed points for homogeneous population under bidirectional multiplicative mutation with shifted payoffs. We have set $\epsilon=4$, $c=5$, and $q=0.22$ for illustrative purpose. We have taken different initial conditions with initial value of $x_1$, $x_1(0)$, set to $0.1$ but different values of $x_2(0)$ (as written in the figure). The asymptotic behaviours of all the timeseries generated clearly belong to two distinct classes.}
	\label{fig:b6}
\end{figure}
\begin{figure*}
	\centering	
	\includegraphics[scale=0.2]{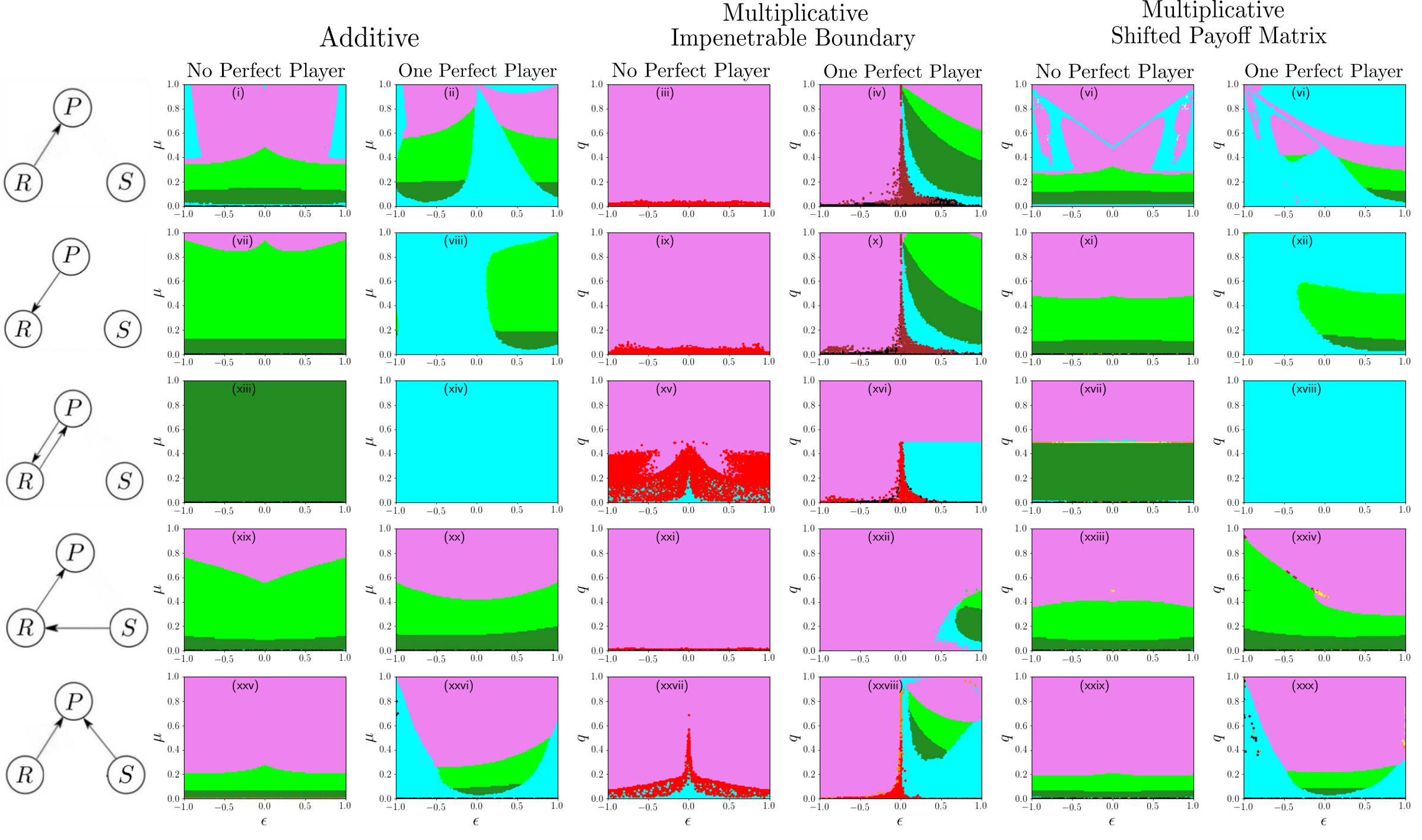}
	\caption{Classification of asymptotic outcomes in $\epsilon$-$\mu$ (and $\epsilon$-$q$) parameter space  under different single and double mutation motifs in non-homogeneous population. The forest-green, the lime, the cyan, the magenta, and the red regions respectively denote stable internal fixed point within the circular neighbourhood of radius 0.1 about the NE, stable internal fixed point but outside the circular neighbourhood of radius 0.1 about the NE, oscillatory limit cycle solution, stable boundary point, and chaotic outcomes. The brown region represent the coexistence of stable internal fixed point and boundary stable fixed point. In the last column of the figure, the shift parameter is fixed as $c=1$.}
	\label{fig:boat6}
\end{figure*}
\begin{figure}
	\centering
	\includegraphics[scale=0.5]{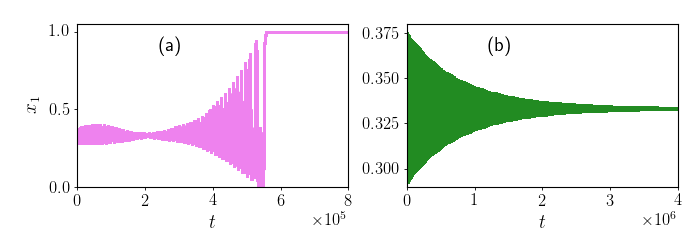}
	\caption{Coexistence of the internal and the boundary fixed points for non-homogeneous population under multiplicative single  mutation motif ($R\to P$) with impenetrable boundary condition. Fixed parameters are $\epsilon=0.1$ and $q=0.1$. The two different initial conditions chosen are (subplot a) $(x_1,x_2,y_1,y_2)=(0.373, 0.315, 0.366, 0.372)$ and (subplot b) $(x_1,x_2,y_1,y_2)=(0.329, 0.301, 0.342, 0.307)$.}
	\label{fig:b7}
\end{figure}

 An interesting conclusion can be made when one player is making mistakes but the other one is perfect---the latter's mutation matrix is identity matrix. The eigenvalues in this case boils down to
\begin{subequations}
	\begin{eqnarray}
	&&\lambda_1^{\pm}=\frac{1}{12}(-9cq-3\epsilon q\pm 2\sqrt{\alpha_2}),\\
	&&\lambda_2^{\pm}=\frac{1}{12}(-3cq-3\epsilon q\pm 2\sqrt{\beta_2}),
	\end{eqnarray}
\end{subequations}
where, $\alpha_2\equiv(-9cq-3\epsilon q)^2/4+2(3+\epsilon^2)(3q-2)$ and
$\beta_2\equiv(-3cq-3\epsilon q)^2/4+2(3+\epsilon^2)(3q-2)$. Quick inspection reveals that for $q<2/3$ the NE is stable and for $q>2/3$ it is unstable. The interesting point is that whether the learning happens for the mistake or mutation probabilities is independent of the shift parameter when one of the players is perfect.

\subsubsection{Impenetrable Boundary}
Next, coming to the case where we do not add a shift parameter to the payoff matrices to achieve forward invariability; rather, we put an impenetrable boundary. Now in presence of mutation matrix $\sf Q$ [see Eq.~(\ref{16})] for both the player and with the payoff matrices Eq.~(\ref{eq:AB}), the NE remains the internal fixed point. Linear stability analysis around the NE point yields repeated eigenvalues,
	\begin{eqnarray}
	\lambda_{\pm}=\pm\sqrt{(1-q)\alpha_3+\beta_3},
		\end{eqnarray}
where $\alpha_3\equiv-6-2\epsilon^2+9q$ and
$\beta_3\equiv(3+4\epsilon^2)q-(9/2)q^2$.
 When the expression inside the square-root is positive, the fixed point is unstable; whereas when the expression is non-positive, the fixed point becomes non-hyperbolic leading to the failure of linear stability analysis. Consequently, in order to predict the asymptotic dynamical outcomes, we resort to numerical exploration. In the former case, trajectories can either settle at a boundary point or become chaotic (see FIG.~\ref{fig:boat2}). Whereas, in the latter case, we have found that the dynamics never seem to converge onto the NE, rather oscillatory outcome or convergence to boundary is seen [see FIG.~\ref{fig:boat1}$(a)$-$(b)$]. In conclusion, neither player learns to play the NE strategy.
 
 This conclusion, however, is bypassed when one of the player turns out to be perfect. In that case, the NE still remains the fixed point and the corresponding eigenvalues of the Jacobian obtained on linearizing the replicator-mutator equation are found to be
 \begin{subequations}
 	\begin{eqnarray}
 	&&\lambda_1^{\pm}=\pm[(1-q)\alpha_3+\beta_3]^{\frac{1}{2}},\\
	&&\lambda_2^{\pm}=\pm\frac{i \sqrt{6+2\epsilon^2}}{3\sqrt{2}},
 	\end{eqnarray}
 \end{subequations} 
Obviously, only in the scenario where $(1-q)\alpha_3+\beta_3\le0$, there is a possibility of learning to happen. We find this numerically [see FIG.~\ref{fig:boat1}$(c)$-$(d)$]: For the illustrative set of parameters, it is clear that when two imperfect players cannot learn, an imperfect player in the presence of a perfect opponent can learn to play the NE.

\subsubsection{Hamiltonian Structure}
Recall that in the absence of mutations in zero-sum RPS game, Hamiltonian chaos is seen. How to find whether a system is Hamiltonian or not has no single fool-proof recipe. We, however, present a case where Hamiltonian structure in the presence of multiplicative mutation can be realized. To appreciate the case, we first must understand that if we add a constant number to all the elements of payoff matrix $\sf A$ and subtract the number from all the  elements of payoff matrix $\sf B$, the replicator equation (in the absence of any mutations) remains invariant. We take this liberty to choose the constant number such that all the elements of matrix $\sf A$ are made positive (of course, all the elements of $\sf B$ become negative).

Now consider the scenario where $q^{x}_{ji}=q^{x}_{j}$ and $q^{y}_{ji}=q^{y}_{j}$ $\forall i$, i.e., the probability of mistake/mutation to a particular strategy is independent of the strategy that mutates. Here Eq.~(\ref{18}) takes the following form: 
\begin{subequations}\label{10}
	\begin{eqnarray}
	&&\dot{x}_1=(q^{x}_{1}-x_1)\mathbf{x}^T \mathbf{{\sf A}y},\\
	&&\dot{x}_2=(q^{x}_{2}-x_2)\mathbf{x}^T \mathbf{{\sf A}y},\\
	&&\dot{y}_1=(q^{y}_{1}-y_1)\mathbf{y}^T \mathbf{{\sf B}x},\\
	&&\dot{y}_2=(q^{y}_{2}-y_2)\mathbf{y}^T \mathbf{{\sf B}x}.
	\end{eqnarray}
\end{subequations}
We realize that for zero-sum bimatrix RPS game, 
\begin{equation}
\mathbf{y}^T \mathbf{{\sf B}x}=\mathbf{x}^T \textsf{B}^T \mathbf{y}=-(\mathbf{x}^T \mathbf{{\sf A}y}),
\end{equation}
facilitating writing of Eq.~(\ref{10}) in compact form as below
\begin{equation}\label{11}
\dot{u}_i=(\mathbf{x}^T \mathbf{{\sf A}y}) {\sf J} \nabla_{\bf u} H,~~~~\forall i \in \{1,2,3,4\}
\end{equation}
where $\mathbf{u}=(u_1,u_2,u_3,u_4)\equiv(x_1,x_2,y_1,y_2)$,
$H=q^{x}_{1}y_1+q^{x}_{2} y_2+q^{y}_{1} x_1+q^{y}_{2} x_2-x_1y_1-x_2y_2$, and ${\sf J}=\tiny{\begin{bmatrix}
0 & {\sf I} \\
-{\sf I} & 0 
\end{bmatrix}}$ ($\sf I$ is $2\times2$ identity matrix). In this case, the positive-definite $(x^T Ay)$, may be absorbed in time (which makes time non-linear) and system (\ref{11}) becomes topologically a Hamiltonian system. 

 The Hamiltonian system has only one fixed point $(q^{x}_{1}, q^{x}_{2}, q^{y}_{1}, q^{y}_{2})$. The linear stability analysis shows that two of the eigenvalues corresponding to $x_1$-axis and $x_2$-axis, the eigendirections, are negative; and other two corresponding to the other two eigendirections $y_1$-axis and $y_2$-axis are positive, i.e., the fixed point behaves as a higher order saddle point making the Hamiltonian system unbounded. This requires us to impose the impenetrable boundary condition in order to achieve the forward-invariability.

Intriguingly, the fixed point is completely specified by the probability of mistake: The NE is fixed point only if $q^x_i=q^y_i=1/3$ $\forall i$. For any other values of mutation parameter, the NE is dynamically unattainable. In fact, since a Hamiltonian system cannot possess any attractor or repeller, the possibility that both the players' strategies asymptotically reach the NE does not exist. Nevertheless, learning is possible for the player one (with strategies $\bf x$), while the other player---owing to the impenetrable boundary condition---can only play one of the pure strategies at equilibrium. This result is illustrated in  FIG.~\ref{fig:boat5}.
 
\section{Further remarks}

\subsection{Other Mutation Motifs}
The analysis of global mutation case has been presented hitherto because of the analytical tractability it renders owing to inherent symmetry of the motif. Other motifs of mutation pattern, as shown in FIG.~\ref{fig:my_label11}, can also be analyzed in similar manner, although now the numerical tools are indispensable.

In the case of homogeneous population, we solve the dynamical equation to find all possible fixed points for a given value of mistake and payoff matrix. Next we calculate the eigenvalues of Jacobian corresponding to the linearized dynamics about those fixed points in order to ascertain the stability of those fixed points. Furthermore, we check the condition of Hopf bifurcation at the internal fixed point to locate any limit-cycle outcome. This way we arrive at different possible asymptotic dynamical outcomes and the possibility of learning in the parameter space, as summarized in FIG.~\ref{fig:table}. It should be pointed out that in the case of multiplicative mistake with impenetrable boundary, although we are able to predict the asymptotic stability of all possible fixed points analytically, the analytic technique is not enough for finding limit-cycles and their stability; thus, we resort to numerics. One observation, in the context of multiplicative mutation with shifted payoff, is that the learning can depend on the choice of initial condition even for a fixed set of parameter values. A specific case is distinctly highlighted in FIG.~\ref{fig:b6}.

In the non-homogeneous scenario, analytically predicting all possible fixed points and their stabilities is challenging. So we have mostly performed numerical analyses. Because our goal is to investigate the learning of NE, we randomly take a few initial conditions close to NE and check its long term the dynamical behaviour as shown in FIG.~\ref{fig:boat6}. The chaotic regions have been double checked using the positivity of corresponding maximum Lyapunov exponent. Here also bistability is witnessed: Under multiplicative single mutation motif with impenetrable boundary condition imposed, coexistence of stable internal and boundary states appear, i.e., learning depends on the choice of initial strategy (illustrated through timeseries in FIG.~\ref{fig:b7}) for a fixed set of parameters. However, the bistability involving enforcement of impenetrable boundary condition could change if the protocol adopted to bound trajectory is changed. We elaborately discuss below the facet of imposing boundary condition in different ways.
\begin{figure*}
	\centering
	\includegraphics[scale=0.18]{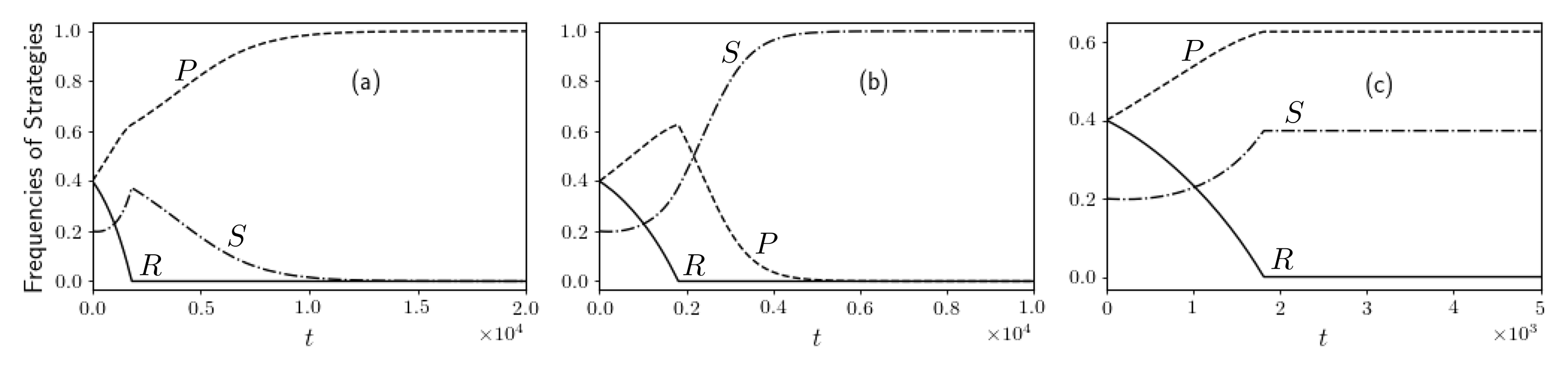}
	\caption{Different protocols at the impenetrable boundary result in distinct evolutionary outcomes in the case of multiplicative mutation: $P\rightarrow R$. Here, $q=0.8$ and $\epsilon=1$.}
	\label{fig:b8}
\end{figure*}
\subsection{Effect of Boundary Conditons}
\label{section_6}
For the sake of simplicity of discussion, let us confine ourselves to the homogeneous population case for the time being; the other case can be analogously understood. First recall that the (non-negative) frequencies of all the strategies must obey the normalization condition: At any instant of time, $\sum_{i=1}^3x_i=1$. When there is no mutation, the dynamics of replicator equation on reaching any of the boundary faces of simplex stays there. In the presence of multiplicative mutation, however, the dynamics of the replicator-mutator equation becomes tricky on the boundaries: The trajectories can now leak out of the simplex if the dynamics is let go ad infinitum. The normailzation condition is still valid but now with negative and greater-than-unity frequencies which are totally unphysical. Therefore, this mathematical problem has to be dealt with in rather ad hoc manner which may be justified physically depending on the particular scenario in hand.

Before we illustrate how different protocols of time-evolution at the boundaries compare, we must emphasize that our assertion about mistakes facilitating learning is independent of the particular protocol adopted at the boundary. Particularly, it is so because learning can happen when the internal fixed point is \emph{locally} asymptotically stable. Our conclusions are based on (local) linear stability analysis about the internal fixed point; this analysis  is completely unaffected by the type of protocol we adopt at the boundaries. Likewise, the criterion on mutation-matrix for the existence of Hamiltonian structure remains intact; only the frequencies of strategies of the player who cannot learn, settle down at different boundary points. 

Coming back to specifying the protocol at boundaries that we have adopted in this paper, let us throw a glance at FIG.~\ref{fig:b8}. The subplot (a) has been generated as follows. There are three boundaries each of which can be uniquely specified by which out of the three frequencies vanish. Consider, for example, the boundary face where the frequency ($x_1$) of $R$ is zero. When orbit reach here, we allow---as we do normally at all times---the time-evolution of $x_1$ and $x_2$ (frequency of $P$) with a view to finding $x_3$ (frequency of $S$) through the normalization condition. Here, if the frequency of $R$ becomes negative (trying to move out of the simplex), we pull it back on the boundary, i.e., we set the frequency to zero. Subsequently, we find $x_2$ using the flow vector component $\dot{x}_2$ and $x_3$ as $1-x_2$. We iterate the process and sometimes, because of mutations, $x_1$ can become more than zero and the trajectory leaves the boundary to reenter the simplex's interior. We follow analogous prescription at the other two boundaries for keeping the dynamics bounded on the simplex.

Because of the freedom of which frequency to find using the corresponding flow vector component and which one to find using the normalization condition, one could have adopted other protocols of keeping the dynamics bounded on the simplex. This however could lead to altogether different long term global dynamics. See, e.g., FIG.~\ref{fig:b8}(b). Here, on the boundary where $x_1=0$, we find $x_3$ using the flow vector component $\dot{x}_3$ and $x_2$ as $1-x_3$. Furthermore, we illustrate in FIG.~\ref{fig:b8}(c) result of yet another protocol where as soon as any one of the strategies reaches the boundary of simplex, we completely cease the evolution of all the strategies.

Which protocol to make the boundaries impenetrable is most appropriate is open to interpretation. Sometimes some protocol can lead to counterintuitive results, e.g., observe that in FIG.~\ref{fig:b8}(a), the frequency of strategies evolve ultimately to only have $P$-strategy in the population. This is quite interesting as although the only mutation considered here is from $P$ to $R$, no $R$-strategists are present in a population with all $P$-strategists.
\section{Discussion}
The folk theorem of evolutionary game theory~\cite{Cressman2014pnas} elevates the relevance of the replicator equation to one of the cornerstones of the theory: A dynamically stable convergent outcome may be read as the game theoretic equilibrium that two competing von Neumann--Morgenstern rational players~\cite{vNM}, with intelligence and common knowledge about certain aspects of the game, would settle at. The theorem is silent about the cases where mutation or mistake are present. Thus, whether mutation or mistake can guide the evolutionary dynamics towards the rational outcome, i.e., Nash equilibrium, is something worth pondering.

A general answer to this question appears even more challenging in the light of the facts that (i) there are many mathematically non-equivalent ways of incorporating mutation in the replicator equation, (ii) the forward invariance of the simplex is not guaranteed automatically in the presence of mutation, and (iii) there can be a lot of possible motifs of mutation among the strategies. Thus, in this paper, we have delved into an extensive case study of evolution of the players playing the RPS game where mistakes or mutations are allowed. While we have mostly presented the study as one for the learning scenario, we have made it crystal clear that the results carry over to biological scenarios as well.

We have been very careful in analytically deriving the replicator-mutator models from  scratch using some well-known learning mechanisms and also in making sure that the parameter values we have worked with render the underlying simplex forward invariant. Subsequently, we have divulged instances of convergent outcomes (fixed point), periodically repeating outcomes (limit cycles), and randomly changing outcomes (chaotic solutions). The most interesting finding is that mistakes can actually lead to learning of the strategies that in turn leads to rational outcome in some cases where the dynamics would otherwise drive the system away from the NE---sometimes due to chaos.

We hope this paper will invite the researchers to expand on the results found herein and generalize them for arbitrary games with any number of strategies; in the course of such investigations, it is hoped that finally one will find generalization of the folk theorem of evolutionary game theory in the presence of mutations and mistakes.
\acknowledgements
Sagar Chakraborty acknowledges the support from SERB (DST, govt. of India) through project no. MTR/2021/000119.

\appendix

\section{Replicator-Mutator Equation in Learning }\label{a1}
There are several mechanisms of reinforcement learning. What we succinctly want to show in this section is that some of the well-known mechanisms do lead to replicator-mutator equations in some asymptotic limit. Let us begin with the Bush--Mosteller learning modified to inclide the effect of mistake.

\subsection{The Bush--Mosteller Learning}\label{app-1}
Consider the classical scenario of the Bush--Mosteller (BM) learning~{\cite{macy1991UCP,FLACHE2002JCR,Macy2002PNAS,borgers2000IER}} where two competing players do not have prior knowledge about the action-payoff correlation. Furthermore, they need not know the payoffs of opponent player and need not have any belief about the opponent’s strategy. Every player, in line with the law of effect~\cite{gray2010psychology}, plays a pure strategy with a frequency that depends on the payoff obtained in the immediately preceding round during the repeated interaction between the players.

We can safely adopt the notations introduced in the main text in the context of two-player bimatrix games along with the following two changes: Firstly, since the number of rounds of play is discrete, $x_i(n)$ (and $y_i(n)$) denote the frequencies of the first (and second) player's $i$th strategy at $n$th round. Secondly, we restrict the payoff elements of $\sf A$ and $\sf B$---respectively denoted by $A_{ij}$ and $B_{ij}$---to lie between zero to unity: This is merely a matter of convenient normalization.

As per the BM learning mechanism, more the payoff in a round, more the frequency of the corresponding strategy in the next round, i.e.,
\begin{subequations}
	\begin{eqnarray}
		&& x_j (n+1)=A_{jk}+(1-A_{jk})x_j(n) \\&&
		x_{i}(n+1)=(1-A_{jk})x_{i}(n) ~\forall i \ne j,
	\end{eqnarray}
\end{subequations}
and likewise for the other player. One could now model the effects of mistakes into it in at least two ways.

One way is when the player one (say) mistakenly chooses a different action with a frequency more than what the BM learning equation directs. It is easy to express this mathematically as follows:
\begin{subequations}\label{eq:bma}
	\begin{eqnarray}
		&& x_j (n+1)=A_{jk}+(1-A_{jk})x_j(n)-\sum_{\substack{i=1\\i\ne j}}^{J}\tilde{\mu}^x_{ij} \\&&
		x_{i}(n+1)=(1-A_{jk})x_{i}(n)+\tilde{\mu}^x_{ij} ~\forall i \ne j.~~~~~~~
	\end{eqnarray}
\end{subequations}
Here  $\tilde{\mu}^x_{ij}$ is the probability of mistake that leads to choosing $i$th strategy when $j$th is supposed to be played. 

Another way could be that a (forgetful) player makes mistake in associating the payoff obtained with the strategy that leads to it. Mathematically,
\begin{subequations}\label{eq:bmm}
	\begin{eqnarray}
		&&x_j(n+1)=A_{jk}q^x_{jj}+(1-A_{jk})x_j \\&&
		x_{i}(n+1)=A_{jk}q^x_{ij}+(1-A_{jk})x_{i}~\forall i \ne j~~~~.
	\end{eqnarray}
\end{subequations}
Here, $q^x_{ij}$ is the probability measure of mistake from $j$ to $i$ strategy of player one.

Now the process of arriving at the deterministic differential equation is straightforward~\cite{Borgers1997JET}: One needs to calculate $\mathbb{E}[\Delta x_j (n)|({\bf x}, {\bf y})]$ where $\Delta x_j(n)\equiv x_{j}(n+1)-x_j(n)$, $\Delta y_j(n)\equiv y_j(n+1)-y_j(n)$, and ${\mathbb{E}}$ denotes the operation of taking expectation over the present state. Subsequently, one introduces a small parameter $\theta$ so as to redefine the payoff matrices as $\theta{\sf A}$ and $\theta{\sf B}$, and additive mutation probability as $\theta\tilde{\mu}^x_{ij}$. Finally, we recognize $\lim_{\theta\to0}\{\mathbb{E}[\Delta x_j (n)|({\bf x}(n), {\bf y}(n))]/\theta\}$ as $dx_i/dt$ to arrive at Eqs.~(\ref{17}) and Eqs.~(\ref{18}).

\subsection{The Roth--Erev Learning}
\label{app-2}
The BM learning does not obey the law of practice~\cite{snoddy1926jap}. However, the Roth--Erev (RE) learning~\cite{Roth1995GEB,erev1998JSTOR} does. The main difference from the BM mechanism is that here the players have memory beyond merely the immediately preceding round. Let us write down the learning process in the setting of a two-player game mathematically---modified to include the instance of mistakes as done above.

When the first player plays strategy $j$ against the second player's strategy $k$, the payoff obtained is $A_{jk}$. One can define prosensity, $p_j^1(n)$, of playing strategy $j$ by the player one at round $n$ as total cummulative payoff obained whenever $j$th strategy has been played till round $n$. If $Q_1(n)=\sum_{j=1}^Jp^1_j(n)$ is the total accumulated payoff (calculted irrespective of which strategy player one implements) obtained till round $n$, then the RE learning process says that 
\begin{subequations}\label{20}
	\begin{eqnarray}
		&&p_{j}^1(n+1)=p_{j}^1(n)+A_{jk},\\&&
		p_{i}^1(n+1)=p_{i}^1(n)~~\forall i \ne j,
	\end{eqnarray}
\end{subequations}
and
\begin{equation*}
	x_j(n)=\frac{p_j^1(n)}{Q_1(n)} ~~\text{and}~~ y_k(n)=\frac{p_k^2(n)}{Q_2(n)},
\end{equation*} 
where the latter relations in aforementioned set of equation can be intepreted analogously for the second player.

Focusing on the effect of the addtive mutation, we observe that the RE learning process should be recast as
\begin{subequations}\label{24}
	\begin{eqnarray}
		&&p_{j}^1(n+1)=p_{j}^1(n)+A_{jk}-\sum_{\substack{i=1\\i\ne j}}^{J}\mu^x_{ij},\\&&
		p_{i}^1(n+1)=p_{i}^1(n)+\mu^x_{ij}~\forall i \ne j,
	\end{eqnarray}
\end{subequations}
for player one. Here recall $\mu^x_{ij}$'s are the rate of mistake. The contributions to $\Delta x_j$ come in two ways: (i) playing strategy $j$ in previous rounds and (ii) playing non-$j$ strategies in previous rounds. The former contributes to both the propensity and total accumulated payoff, while the latter contributes only to the total accumulated payoff. These two contributions can respectively be found as
\begin{eqnarray}
	\Delta x_j=\frac{(1-x_{j}(n))A_{jk}-\sum_{\substack{i=1\\i\ne j}}^{J}\mu^x_{ij}}{Q_1(n)}+\mathcal{O}\left(\frac{1}{Q_1^2(n)}\right),~~~~
\end{eqnarray}
and 
\begin{equation}
	\Delta x_j=\frac{-(x_{j}(n))A_{ik}+\mu^x_{ji}}{Q_1(n)}+\mathcal{O}\left(\frac{1}{Q_1^2(n)}\right).
\end{equation}
Interpreting $\lim_{n\to\infty}\{\mathbb{E}[\Delta x_j (n)|({\bf x}(n), {\bf y}(n))]/Q_1^{-1}(n)\}$ as $dx_i/dt$, while requiring~\cite{hopkins2002econometrica} $\lim\limits_{n \to \infty}Q_1(n)/Q_2(n) \approx 1$,  we effectively derive Eqs.~(\ref{17}).

The derivation of the replicator-mutator equation with multiplicative mutation is along the similar lines. The RE learning process first needs to be modified to
\begin{subequations}\label{21}
	\begin{eqnarray}
		&&p_{j}^1(n+1)=p_{j}^1(n)+q^x_{jj}A_{jk},\\&&
		p_{i}^1(n+1)=p_{i}^1(n)+q^x_{ij}A_{jk}~~\forall i \ne j.
	\end{eqnarray}
\end{subequations}
Subsequently, the expectation value of $\Delta x_j$ and $\Delta y_j$ has to be found and then taking the limits (as invoked in the additive case) yields Eqs.~(\ref{18}).

\begin{table*}	
	\begin{center}
		\begin{tabular}{ |c|c|c|c|c| } 
			\hline
			Focal player's strategy & {Opponent player's strategy } & 	$Pr(T_1)$ & $Pr(T_2)$ \\
			\hline
			{$T_1$} & $T_1$ & $1-\tilde{\mu}_{21}$ & $\tilde{\mu}_{21}$\\ 
			$T_1$ & $T_2$ & $\frac{1}{2}+\beta \{(\sf \Pi \textbf{x})_1-(\sf \Pi \textbf{x})_2\}-\tilde{\mu}_{21}$ &  $\frac{1}{2}+\beta \{(\sf \Pi \textbf{x})_2-(\sf \Pi \textbf{x})_1\}+\tilde{\mu}_{21}$\\
			$T_2$ & $T_1$ & $\frac{1}{2}+\beta \{(\sf \Pi \textbf{x})_1-(\sf \Pi \textbf{x})_2\}+\tilde{\mu}_{12}$ &  $\frac{1}{2}+\beta \{(\sf \Pi \textbf{x})_2-(\sf \Pi \textbf{x})_1\}-\tilde{\mu}_{12}$\\ 
			$T_2$ & $T_2$ & $\tilde{\mu}_{12}$ & $1-\tilde{\mu}_{12}$\\ 
			\hline
		\end{tabular}
		\caption{\label{table_1} Updation rules in the social learning with additive mistakes.}
	\end{center}
\end{table*}
\begin{table*}	
	\begin{center}
		\begin{tabular}{ |c|c|c|c|c| } 
			\hline
			Focal player's strategy &  {Opponent player's strategy} & $Pr(T_1)$ & $Pr(T_2)$ \\
			\hline
			{$T_1$} & $T_1$ & $1+\beta \big\{(1-\tilde{q}_{21})\cdot(\sf \Pi \textbf{x})_1-(\sf \Pi \textbf{x})_1\big\}$ & $\beta \big\{({\sf \Pi} \textbf{x})_1-(1-\tilde{q}_{21})\cdot(\sf \Pi \textbf{x})_1\big\}$\\ 
			$T_1$ & $T_2$ & $\frac{1}{2}+\beta \big\{(1-\tilde{q}_{21})\cdot(\sf \Pi \textbf{x})_1-(\sf \Pi \textbf{x})_2\big\}$ &  $\frac{1}{2}+\beta \big\{({\sf \Pi} \textbf{x})_2-(1-\tilde{q}_{21})\cdot(\sf \Pi \textbf{x})_1\big\}$\\ 
			$T_2$ & $T_1$ & $\frac{1}{2}+\beta \big\{({\sf \Pi} \textbf{x})_1-(1-\tilde{q}_{12})\cdot(\sf \Pi \textbf{x})_2\big\}$  &  $\frac{1}{2}+\beta \big\{(1-\tilde{q}_{12})\cdot(\sf \Pi \textbf{x})_2-(\sf \Pi \textbf{x})_1\big\}$\\ 
			$T_2$ & $T_2$ & $\beta \big\{({\sf \Pi} \textbf{x})_2-(1-\tilde{q}_{12})\cdot(\sf \Pi \textbf{x})_2\big\}$ & $1+\beta \big\{(1-\tilde{q}_{12})\cdot(\sf \Pi \textbf{x})_2-(\sf \Pi \textbf{x})_2\big\}$\\ 
			\hline
		\end{tabular}
		\caption{\label{table_2} Updation rules in the social learning with multiplicative mistakes.}
	\end{center}
\end{table*}
\subsection{The Social Learning}\label{app-3}

In a homogeneous, well-mixed, and infinite-sized population, the  mathematical model of the social learning~{\cite{mcelreath2008book}} is easily amenable to inclusion of mistake or mutation. Let us consider a population consisting of only two types of individuals, $T_1$ and $T_2$ with frequencies  $x_1=x$ and $x_2=1-x$ respectively. If the payoff matrix is $\sf \Pi$, then the expected payoffs of type $T_1$ and $T_2$ are $({\sf \Pi} {\bf x})_1$ and $({\sf \Pi} {\bf x})_2$ respectively on playing a one-shot game. After getting her payoff, a player compares it with that of another randomly chosen individual (who can be either $T_1$ or $T_2$) and update her type. The updation rules are given in TABLE~\ref{table_1} {and TABLE~\ref{table_2} respectively for the cases of additive and multiplicative mutations. Here $Pr(T_1)$ and $Pr(T_2)$ is the probabilities of choosing the strategy $T_1$ and $T_2$, respectively}. $\beta$ is a real parameter quantifying the selection strength. 

{TABLE~\ref{table_1} includes the effect of the additive mistake: For instance, an individual of type $T_1$ ($T_2$) can mistakenly choose action $T_2$ ($T_1$) with probability $\tilde {\mu}_{21}$ ($\tilde {\mu}_{12}$). On the other hand, TABLE~\ref{table_2} includes the effect of the multiplicative mistake: A forgetful individual of type $T_1$ ($T_2$) can mistakenly associate her payoff to action $T_2$ ($T_1$) with probability $\tilde{q}_{21}$ ($\tilde{q}_{12}$).} Note that when the expected payoffs of $T_1$ and $T_2$ are same and there are no mistakes, then with probability half, players can choose any one of the strategies. 

{Now the expected change in frequency of $T_1$ type of individuals in the presence of additive mistake is}
\begin{eqnarray}
&&\mathbb{E}[\Delta x|{\bf x}]
=x^2\cdot(1-\tilde{\mu}_{21})\nonumber\\
&&\phantom{\mathbb{E}[\Delta x|{\bf x}]}+x(1-x)\cdot\left[\frac{1}{2}+\beta \{(\sf \Pi \textbf{x})_1-(\sf \Pi \textbf{x})_2\}-\tilde{\mu}_{21}\right]\nonumber\\
&&\phantom{\mathbb{E}[\Delta x|{\bf x}]}+x(1-x)\cdot\left[\frac{1}{2}+\beta \{(\sf \Pi \textbf{x})_1-(\sf \Pi \textbf{x})_2\}+\tilde{\mu}_{12}\right]\nonumber\\
&&\phantom{\mathbb{E}[\Delta x|{\bf x}]}+(1-x)^2\cdot\tilde{\mu}_{12}-x,
\end{eqnarray}
which ultimately simplifies to the following expression:
\begin{eqnarray}
	&&\mathbb{E}[\Delta x|{\bf x}]=2\beta x(1-x)\left[(\sf \Pi \textbf{x})_1-(\sf \Pi \textbf{x})_2\right]\nonumber\\
	&&\phantom{\mathbb{E}[\Delta x|{\bf x}]=}+\tilde{\mu}_{12} (1-x)-\tilde{\mu}_{21}x.
\end{eqnarray}
Introducing the rate of making mistakes, ${\mu_{12}\equiv  \tilde{\mu}_{12}/2\beta}$ and ${\mu_{21}\equiv \tilde{\mu}_{21}/2\beta}$, and identifying $\lim_{\beta\to0}\{\mathbb{E}[\Delta x|{\bf x}]/2\beta\}$ with $dx/dt$, we derive Eqs.~(\ref{6}) for two types of players.

{ Similarly, the expected change in frequency of $T_1$ type of individuals in the presence of multiplicative mistake is	
\begin{eqnarray}
&&\mathbb{E}[\Delta x|{\bf x}]
=x^2\cdot \bigg[1+\beta \big\{(1-\tilde{q}_{21})\cdot({\sf \Pi} \textbf{x})_1-({\sf \Pi} \textbf{x})_1\big\}\bigg]\nonumber\\&& \phantom{\mathbb{E}[\Delta x|{\bf x}]}
+x(1-x)\cdot\left[\frac{1}{2}+\beta \big\{(1-\tilde{q}_{21})\cdot(\sf \Pi \textbf{x})_1-(\sf \Pi \textbf{x})_2\big\}\right]\nonumber\\
&&\phantom{\mathbb{E}[\Delta x|{\bf x}]}+x(1-x)\cdot\left[\frac{1}{2}+\beta \big\{({\sf \Pi} \textbf{x})_1-(1-\tilde{q}_{12})\cdot(\sf \Pi \textbf{x})_2\big\}\right]\nonumber\\&&\phantom{\mathbb{E}[\Delta x|{\bf x}]}+(1-x)^2\cdot \bigg[\beta \big\{({\sf \Pi} \textbf{x})_2-(1-\tilde{q}_{12})\cdot({\sf \Pi} \textbf{x})_2\big\}\bigg]\nonumber\\&&\phantom{\mathbb{E}[\Delta x|{\bf x}]}-x,
\end{eqnarray}
which ultimately simplifies to the following expression
\begin{eqnarray}
&&\mathbb{E}[\Delta x|{\bf x}]=2\beta x(1-x)\left[(\sf \Pi \textbf{x})_1-(\sf \Pi \textbf{x})_2\right]\nonumber\\
&&\phantom{\mathbb{E}[\Delta x|{\bf x}]=}-\beta[\tilde{q}_{21}x({\sf \Pi} \textbf{x})_1-\tilde{q}_{12}(1-x)(\sf \Pi \textbf{x})_2].~~~~~~
\end{eqnarray}
One can rewrite above expression in more compact form as written below,
\begin{eqnarray}
&&\mathbb{E}[\Delta x|{\bf x}]={2\beta} \left\{\frac{(2-\tilde{q}_{21})}{2}x(\mathbf{{\sf \Pi}x})_1+\frac{\tilde{q}_{12}}{2}(1-x)(\mathbf{{\sf \Pi}x})_2\right\}\nonumber\\&&\phantom{\mathbb{E}[\Delta x|{\bf x}]=}-{2\beta}x \cdot({\bf x}^T\mathbf{{\sf \Pi}x}).
\end{eqnarray}
In order to reach the form of replicator-mutator equation (for two types of players), given in Eqs.~(\ref{7}), it is convenient to redefine the probabilities of mistake, $q_{21} \equiv \tilde{q}_{21}/2$ and $q_{12} \equiv \tilde{q}_{12}/2$. Subsequently, as before, we identify $\lim_{\beta\to0}\{\mathbb{E}[\Delta x|{\bf x}]/2\beta\}$ with $dx/dt$ to arrive at the desired equation. 

The extension of derivations presented in this section is straightforward for homogeneous population with $J$ types of individuals.}
\section{Forward Invariance}
\label{app:fi}
The simplex on which the replicator-mutator equation is defined should ideally be forward-invariant: Any point outside the simplex is not a physically meaningful state. For the case of the replicator equation (without any mutation incoorporated), the forward-invariance is trivially satisfied; however, this is not always guaranteed with mutation is modeled into it. One may need to restrict the range of parameters' values. For the sake of clarity, we present this for the case of homogeneous population for which Eqs.~(\ref{6})~and~(\ref{7}) were considered. 

Let us consider $J-1$ dimensional simplex $S_J=\{(x_1,x_2,\cdots,x_J):x_1+x_2+\cdots+x_J=1,\,x_i\in[0,1] \,\forall i\in\{1,2,\cdots,J\}\}$ embedded in ${\mathbb R}^J$ spanned by the $J$ unit vectors $\hat{\bf e}_i$'s. The simplex remains invariant under the replicator-mutator flow  if and only if, at every boundary point, the flow-vector's component (lying on the simplex) which is perpendicular to the boundary at that point is towards the interior of the simplex.  Without any loss of generality, consider a boundary surface, $x_i=0$, which corresponds to the following hyperplane: $x_1+x_2+\cdots+x_{i-1}+x_{i+1}+\cdots+x_J=1$, denoted by $(\partial S_J)_i$, say. The (non-normalized) vector that is perpendicular to $(\partial S_J)_i$ has the form ${\bf v}=\hat{\bf e}_1+\hat{\bf e}_2+\cdots+\hat{\bf e}_{i-1}+a \hat{\bf e}_i+\hat{\bf e}_{i+1}+\cdots+\hat{\bf e}_J$, where $\hat{\bf  e}_i$ is an unit vector along the $x_i$-axis of ${\mathbb R}^J$ and $a$ is any real-valued scalar. Presence of $a$ basically says that there are uncountably infinite number of vectors that are perpendicular to $(\partial S_J)_i$. 

We want the specific value of $a$ for which the vector ${\bf v}$ completely lies on simplex $S_J$ because the physical dynamics is confined on the simplex. In other words, the dot product of ${\bf v}$ with the vector normal to $S_J$  must vanish---a condition that makes $a=-(J-1)$, which is always negative. We note that this special $\bf v$ is directed from $(J-1)\hat{\bf e}_i$ to $\hat{\bf e}_1+\hat{\bf e}_2+\cdots+\hat{\bf e}_{i-1}+\hat{\bf e}_{i+1}+\cdots+\hat{\bf e}_J$, i.e., along the outward direction from the simplex's interior. Therefore, the component of the flow vector $\dot{\bf x}$ at  $(\partial S_J)_i$ is given by
\begin{eqnarray}
	&&\dot{x}_1+\dot{x}_2+\cdots+\dot{x}_{i-1}-(J-1)\dot{x}_i+\dot{x}_{i+1}+\cdots+\dot{x}_J\nonumber\quad\\
	&&\phantom{xxxxxxxxxxxxxxxxxxx}=-(J-1)\sum_{\substack{j=1\\j\ne i}}^{J}\mu_{ij}x_j\quad \label{eq:fia}
\end{eqnarray}
and
\begin{eqnarray}
	&&\dot{x}_1+\dot{x}_2+\cdots+\dot{x}_{i-1}-(J-1)\dot{x}_i+\dot{x}_{i+1}+\cdots+\dot{x}_J\nonumber\qquad\\
	&&\phantom{xxxxxxxxxxxxxxxx}=-(J-1)\sum_{\substack{j=1\\j\ne i}}^{J}q_{ij}x_j(Ax)_j\qquad \label{eq:fim}
\end{eqnarray}
respectively for additive and multiplicative mutations.

In Eq.~(\ref{eq:fia}), we see that the R.H.S. is always negative rendering the replicator-mutator flow to be confined within the simplex: The forward-invariability is always guaranteed. However, similar conclusion cannot be drawn for the multiplicative mutation case as the sign of the R.H.S. of Eq.~(\ref{eq:fim}) is payoff matrix dependent: If the payoff matrix element has not negative elements, then for all ${\bf x}$, the R.H.S. is negative; and forward invariability for $S_J$ is achieved. We remark that these conclusions straightforwardly carry over to the nonhomogeneous population case where the phase space is the Cartesian product of two simplices.  
\bibliography{chakraborty_etal_references}

\begin{thebibliography}{61}
\expandafter\ifx\csname natexlab\endcsname\relax\def\natexlab#1{#1}\fi
\expandafter\ifx\csname bibnamefont\endcsname\relax
  \def\bibnamefont#1{#1}\fi
\expandafter\ifx\csname bibfnamefont\endcsname\relax
  \def\bibfnamefont#1{#1}\fi
\expandafter\ifx\csname citenamefont\endcsname\relax
  \def\citenamefont#1{#1}\fi
\expandafter\ifx\csname url\endcsname\relax
  \def\url#1{\texttt{#1}}\fi
\expandafter\ifx\csname urlprefix\endcsname\relax\def\urlprefix{URL }\fi
\providecommand{\bibinfo}[2]{#2}
\providecommand{\eprint}[2][]{\url{#2}}

\bibitem[{\citenamefont{Gray}(2010)}]{gray2010psychology}
\bibinfo{author}{\bibfnamefont{P.}~\bibnamefont{Gray}},
  \emph{\bibinfo{title}{Psychology}}, International edition
  (\bibinfo{publisher}{Worth, New York}, \bibinfo{year}{2010}), ISBN
  \bibinfo{isbn}{9781429252454}.

\bibitem[{\citenamefont{Snoddy}(1926)}]{snoddy1926jap}
\bibinfo{author}{\bibfnamefont{G.~S.} \bibnamefont{Snoddy}},
  \bibinfo{journal}{J. Appl. Psychol.} \textbf{\bibinfo{volume}{10}},
  \bibinfo{pages}{1} (\bibinfo{year}{1926}).

\bibitem[{\citenamefont{Cheney and Seyfarth}(2018)}]{cheney2018book}
\bibinfo{author}{\bibfnamefont{D.~L.} \bibnamefont{Cheney}} \bibnamefont{and}
  \bibinfo{author}{\bibfnamefont{R.~M.} \bibnamefont{Seyfarth}},
  \emph{\bibinfo{title}{How monkeys see the world: Inside the mind of another
  species}} (\bibinfo{publisher}{University of Chicago Press, Chicago},
  \bibinfo{year}{2018}).

\bibitem[{\citenamefont{Dong et~al.}(2023)\citenamefont{Dong, Lin, Nieh, and
  Tan}}]{Dong2023science}
\bibinfo{author}{\bibfnamefont{S.}~\bibnamefont{Dong}},
  \bibinfo{author}{\bibfnamefont{T.}~\bibnamefont{Lin}},
  \bibinfo{author}{\bibfnamefont{J.~C.} \bibnamefont{Nieh}}, \bibnamefont{and}
  \bibinfo{author}{\bibfnamefont{K.}~\bibnamefont{Tan}},
  \bibinfo{journal}{Science} \textbf{\bibinfo{volume}{379}},
  \bibinfo{pages}{1015} (\bibinfo{year}{2023}).

\bibitem[{\citenamefont{Charrier and Sturdy}(2005)}]{charrier2005BP}
\bibinfo{author}{\bibfnamefont{I.}~\bibnamefont{Charrier}} \bibnamefont{and}
  \bibinfo{author}{\bibfnamefont{C.~B.} \bibnamefont{Sturdy}},
  \bibinfo{journal}{Behav. Processes} \textbf{\bibinfo{volume}{70}},
  \bibinfo{pages}{271} (\bibinfo{year}{2005}).

\bibitem[{\citenamefont{Taga and Bassler}(2003)}]{Taga2003PNAS}
\bibinfo{author}{\bibfnamefont{M.~E.} \bibnamefont{Taga}} \bibnamefont{and}
  \bibinfo{author}{\bibfnamefont{B.~L.} \bibnamefont{Bassler}},
  \bibinfo{journal}{Proc. Natl. Acad. Sci.} \textbf{\bibinfo{volume}{100}},
  \bibinfo{pages}{14549} (\bibinfo{year}{2003}).

\bibitem[{\citenamefont{Friedman and Sinervo}(2016)}]{friedman2016oup}
\bibinfo{author}{\bibfnamefont{D.}~\bibnamefont{Friedman}} \bibnamefont{and}
  \bibinfo{author}{\bibfnamefont{B.}~\bibnamefont{Sinervo}},
  \emph{\bibinfo{title}{Evolutionary games in natural, social, and virtual
  worlds}} (\bibinfo{publisher}{Oxford University Press, Oxford},
  \bibinfo{year}{2016}).

\bibitem[{\citenamefont{Roth and Erev}(1995)}]{Roth1995GEB}
\bibinfo{author}{\bibfnamefont{A.~E.} \bibnamefont{Roth}} \bibnamefont{and}
  \bibinfo{author}{\bibfnamefont{I.}~\bibnamefont{Erev}},
  \bibinfo{journal}{Games Econ. Behav.} \textbf{\bibinfo{volume}{8}},
  \bibinfo{pages}{164} (\bibinfo{year}{1995}).

\bibitem[{\citenamefont{Erev and Roth}(1998)}]{erev1998JSTOR}
\bibinfo{author}{\bibfnamefont{I.}~\bibnamefont{Erev}} \bibnamefont{and}
  \bibinfo{author}{\bibfnamefont{A.~E.} \bibnamefont{Roth}},
  \bibinfo{journal}{Am. Econ. Rev.} pp. \bibinfo{pages}{848--881}
  (\bibinfo{year}{1998}).

\bibitem[{\citenamefont{Bush and Mosteller}(1955)}]{Bush1955book}
\bibinfo{author}{\bibfnamefont{R.~R.} \bibnamefont{Bush}} \bibnamefont{and}
  \bibinfo{author}{\bibfnamefont{F.}~\bibnamefont{Mosteller}},
  \emph{\bibinfo{title}{Stochastic models for learning.}}
  (\bibinfo{publisher}{John Wiley {\&} Sons Inc}, \bibinfo{year}{1955}).

\bibitem[{\citenamefont{Fudenberg et~al.}(1998)\citenamefont{Fudenberg, Drew,
  Levine, and Levine}}]{fudenberg1998book}
\bibinfo{author}{\bibfnamefont{D.}~\bibnamefont{Fudenberg}},
  \bibinfo{author}{\bibfnamefont{F.}~\bibnamefont{Drew}},
  \bibinfo{author}{\bibfnamefont{D.~K.} \bibnamefont{Levine}},
  \bibnamefont{and} \bibinfo{author}{\bibfnamefont{D.~K.}
  \bibnamefont{Levine}}, \emph{\bibinfo{title}{The theory of learning in
  games}}, vol.~\bibinfo{volume}{2} (\bibinfo{publisher}{MIT press, Cambridge},
  \bibinfo{year}{1998}).

\bibitem[{\citenamefont{McElreath and Boyd}(2008)}]{mcelreath2008book}
\bibinfo{author}{\bibfnamefont{R.}~\bibnamefont{McElreath}} \bibnamefont{and}
  \bibinfo{author}{\bibfnamefont{R.}~\bibnamefont{Boyd}},
  \emph{\bibinfo{title}{Mathematical models of social evolution: A guide for
  the perplexed}} (\bibinfo{publisher}{University of Chicago Press, Chicago},
  \bibinfo{year}{2008}).

\bibitem[{\citenamefont{Taylor and Jonker}(1978)}]{Taylor1978mathbio}
\bibinfo{author}{\bibfnamefont{P.~D.} \bibnamefont{Taylor}} \bibnamefont{and}
  \bibinfo{author}{\bibfnamefont{L.~B.} \bibnamefont{Jonker}},
  \bibinfo{journal}{Math. Biosci.} \textbf{\bibinfo{volume}{40}},
  \bibinfo{pages}{145} (\bibinfo{year}{1978}).

\bibitem[{Bor(1997)}]{Borgers1997JET}
 (\bibinfo{year}{1997}).

\bibitem[{\citenamefont{Hopkins and Posch}(2005)}]{Hopkins2005GEB}
\bibinfo{author}{\bibfnamefont{E.}~\bibnamefont{Hopkins}} \bibnamefont{and}
  \bibinfo{author}{\bibfnamefont{M.}~\bibnamefont{Posch}},
  \bibinfo{journal}{Games Econ. Behav.} \textbf{\bibinfo{volume}{53}},
  \bibinfo{pages}{110} (\bibinfo{year}{2005}).

\bibitem[{\citenamefont{Beggs}(2005)}]{Beggs2005JET}
\bibinfo{author}{\bibfnamefont{A.}~\bibnamefont{Beggs}}, \bibinfo{journal}{J.
  Econ. Theory} \textbf{\bibinfo{volume}{122}}, \bibinfo{pages}{1}
  (\bibinfo{year}{2005}).

\bibitem[{\citenamefont{Maynard~Smith and Price}(1973)}]{SMITH1973nature}
\bibinfo{author}{\bibfnamefont{J.}~\bibnamefont{Maynard~Smith}}
  \bibnamefont{and} \bibinfo{author}{\bibfnamefont{G.~R.} \bibnamefont{Price}},
  \bibinfo{journal}{Nature} \textbf{\bibinfo{volume}{246}}, \bibinfo{pages}{15}
  (\bibinfo{year}{1973}).

\bibitem[{\citenamefont{Maynard~Smith}(1993)}]{smith1993theory}
\bibinfo{author}{\bibfnamefont{J.}~\bibnamefont{Maynard~Smith}},
  \emph{\bibinfo{title}{The theory of evolution}}
  (\bibinfo{publisher}{Cambridge University Press, Cambridge},
  \bibinfo{year}{1993}).

\bibitem[{\citenamefont{Harper}(2009)}]{harper2009arxiv}
\bibinfo{author}{\bibfnamefont{M.}~\bibnamefont{Harper}},
  \emph{\bibinfo{title}{The replicator equation as an inference dynamic}}
  (\bibinfo{year}{2009}).

\bibitem[{\citenamefont{Evans and Raine}(2014)}]{Evans2014JCPA}
\bibinfo{author}{\bibfnamefont{L.~J.} \bibnamefont{Evans}} \bibnamefont{and}
  \bibinfo{author}{\bibfnamefont{N.~E.} \bibnamefont{Raine}},
  \bibinfo{journal}{J. Comp. Physiol. A} \textbf{\bibinfo{volume}{200}},
  \bibinfo{pages}{475} (\bibinfo{year}{2014}).

\bibitem[{\citenamefont{Tulis et~al.}(2016)\citenamefont{Tulis, Steuer, and
  Dresel}}]{Tulis2016FLR}
\bibinfo{author}{\bibfnamefont{M.}~\bibnamefont{Tulis}},
  \bibinfo{author}{\bibfnamefont{G.}~\bibnamefont{Steuer}}, \bibnamefont{and}
  \bibinfo{author}{\bibfnamefont{M.}~\bibnamefont{Dresel}},
  \bibinfo{journal}{Frontline Learn. Res} \textbf{\bibinfo{volume}{4}},
  \bibinfo{pages}{12} (\bibinfo{year}{2016}).

\bibitem[{\citenamefont{Cyr and Anderson}(2015)}]{Cyr2015JEPLMC}
\bibinfo{author}{\bibfnamefont{A.-A.} \bibnamefont{Cyr}} \bibnamefont{and}
  \bibinfo{author}{\bibfnamefont{N.~D.} \bibnamefont{Anderson}},
  \bibinfo{journal}{J. Exp. Psychol. Learn Mem. Cogn.}
  \textbf{\bibinfo{volume}{41}}, \bibinfo{pages}{841} (\bibinfo{year}{2015}).

\bibitem[{\citenamefont{VanLehn}(1988)}]{vanlehn1988book}
\bibinfo{author}{\bibfnamefont{K.}~\bibnamefont{VanLehn}},
  \emph{\bibinfo{title}{Toward a theory of impasse-driven learning}}
  (\bibinfo{publisher}{Springer, New York City}, \bibinfo{year}{1988}).

\bibitem[{\citenamefont{Metcalfe}(2017)}]{Metcalfe2017ARP}
\bibinfo{author}{\bibfnamefont{J.}~\bibnamefont{Metcalfe}},
  \bibinfo{journal}{Annu. Rev. Psychol.} \textbf{\bibinfo{volume}{68}},
  \bibinfo{pages}{465} (\bibinfo{year}{2017}).

\bibitem[{\citenamefont{Stevenson and Stigler}(1994)}]{stevenson1994book}
\bibinfo{author}{\bibfnamefont{H.}~\bibnamefont{Stevenson}} \bibnamefont{and}
  \bibinfo{author}{\bibfnamefont{J.~W.} \bibnamefont{Stigler}},
  \emph{\bibinfo{title}{Learning gap: Why our schools are failing and what we
  can learn from Japanese and Chinese educ}} (\bibinfo{publisher}{Simon and
  Schuster, New York City}, \bibinfo{year}{1994}).

\bibitem[{\citenamefont{Nowak et~al.}(2001)\citenamefont{Nowak, Komarova, and
  Niyogi}}]{Nowak2001science}
\bibinfo{author}{\bibfnamefont{M.~A.} \bibnamefont{Nowak}},
  \bibinfo{author}{\bibfnamefont{N.~L.} \bibnamefont{Komarova}},
  \bibnamefont{and} \bibinfo{author}{\bibfnamefont{P.}~\bibnamefont{Niyogi}},
  \bibinfo{journal}{Science} \textbf{\bibinfo{volume}{291}},
  \bibinfo{pages}{114} (\bibinfo{year}{2001}).

\bibitem[{\citenamefont{Komarova et~al.}(2001)\citenamefont{Komarova, Niyogi,
  and Nowak}}]{KOMAROVA2001jtb}
\bibinfo{author}{\bibfnamefont{N.~L.} \bibnamefont{Komarova}},
  \bibinfo{author}{\bibfnamefont{P.}~\bibnamefont{Niyogi}}, \bibnamefont{and}
  \bibinfo{author}{\bibfnamefont{M.~A.} \bibnamefont{Nowak}},
  \bibinfo{journal}{J. Theor. Biol.} \textbf{\bibinfo{volume}{209}},
  \bibinfo{pages}{43} (\bibinfo{year}{2001}).

\bibitem[{\citenamefont{Mobilia}(2010)}]{Mobilia2010jtb}
\bibinfo{author}{\bibfnamefont{M.}~\bibnamefont{Mobilia}}, \bibinfo{journal}{J.
  Theor. Biol.} \textbf{\bibinfo{volume}{264}}, \bibinfo{pages}{1}
  (\bibinfo{year}{2010}).

\bibitem[{\citenamefont{Pais and Leonard}(2011)}]{Pais2011ieee}
\bibinfo{author}{\bibfnamefont{D.}~\bibnamefont{Pais}} \bibnamefont{and}
  \bibinfo{author}{\bibfnamefont{N.~E.} \bibnamefont{Leonard}}, in
  \emph{\bibinfo{booktitle}{{IEEE} Conference on Decision and Control and
  European Control Conference}} (\bibinfo{publisher}{{IEEE}},
  \bibinfo{year}{2011}).

\bibitem[{\citenamefont{Toupo and Strogatz}(2015)}]{Toupo2015PRE}
\bibinfo{author}{\bibfnamefont{D.~F.~P.} \bibnamefont{Toupo}} \bibnamefont{and}
  \bibinfo{author}{\bibfnamefont{S.~H.} \bibnamefont{Strogatz}},
  \bibinfo{journal}{Phys. Rev. E} \textbf{\bibinfo{volume}{91}},
  \bibinfo{pages}{052907} (\bibinfo{year}{2015}).

\bibitem[{\citenamefont{You et~al.}(2017)\citenamefont{You, Kwon, Jo, Jung, and
  Baek}}]{You2017pre}
\bibinfo{author}{\bibfnamefont{T.}~\bibnamefont{You}},
  \bibinfo{author}{\bibfnamefont{M.}~\bibnamefont{Kwon}},
  \bibinfo{author}{\bibfnamefont{H.-H.} \bibnamefont{Jo}},
  \bibinfo{author}{\bibfnamefont{W.-S.} \bibnamefont{Jung}}, \bibnamefont{and}
  \bibinfo{author}{\bibfnamefont{S.~K.} \bibnamefont{Baek}},
  \bibinfo{journal}{Phys. Rev. E} \textbf{\bibinfo{volume}{96}},
  \bibinfo{pages}{062310} (\bibinfo{year}{2017}).

\bibitem[{\citenamefont{Mittal et~al.}(2020)\citenamefont{Mittal, Mukhopadhyay,
  and Chakraborty}}]{Mittal2020pre}
\bibinfo{author}{\bibfnamefont{S.}~\bibnamefont{Mittal}},
  \bibinfo{author}{\bibfnamefont{A.}~\bibnamefont{Mukhopadhyay}},
  \bibnamefont{and}
  \bibinfo{author}{\bibfnamefont{S.}~\bibnamefont{Chakraborty}},
  \bibinfo{journal}{Phys. Rev. E} \textbf{\bibinfo{volume}{101}},
  \bibinfo{pages}{042410} (\bibinfo{year}{2020}).

\bibitem[{\citenamefont{Mukhopadhyay et~al.}(2021)\citenamefont{Mukhopadhyay,
  Chakraborty, and Chakraborty}}]{Mukhopadhyay2021JoPC}
\bibinfo{author}{\bibfnamefont{A.}~\bibnamefont{Mukhopadhyay}},
  \bibinfo{author}{\bibfnamefont{S.}~\bibnamefont{Chakraborty}},
  \bibnamefont{and}
  \bibinfo{author}{\bibfnamefont{S.}~\bibnamefont{Chakraborty}},
  \bibinfo{journal}{J. phys. Complex} \textbf{\bibinfo{volume}{2}},
  \bibinfo{pages}{035005} (\bibinfo{year}{2021}).

\bibitem[{\citenamefont{Nowak}(2006)}]{nowak2006evolutionary}
\bibinfo{author}{\bibfnamefont{M.~A.} \bibnamefont{Nowak}},
  \emph{\bibinfo{title}{Evolutionary dynamics: exploring the equations of
  life}} (\bibinfo{publisher}{Harvard university press, Cambridge},
  \bibinfo{year}{2006}).

\bibitem[{\citenamefont{Rice}(2004)}]{rice2004evolutionary}
\bibinfo{author}{\bibfnamefont{S.}~\bibnamefont{Rice}},
  \emph{\bibinfo{title}{Evolutionary theory: mathematical and conceptual
  foundations}} (\bibinfo{publisher}{Massachusetts: Sinauer Associates, Inc.},
  \bibinfo{year}{2004}).

\bibitem[{\citenamefont{Walker and Walker}(2004)}]{walker2004ss}
\bibinfo{author}{\bibfnamefont{D.}~\bibnamefont{Walker}} \bibnamefont{and}
  \bibinfo{author}{\bibfnamefont{G.}~\bibnamefont{Walker}},
  \emph{\bibinfo{title}{The official rock paper scissors strategy guide}}
  (\bibinfo{publisher}{Simon and Schuster, New York City},
  \bibinfo{year}{2004}).

\bibitem[{\citenamefont{Livingstone and Nisbet}(2008)}]{livingstone2008cup}
\bibinfo{author}{\bibfnamefont{N.}~\bibnamefont{Livingstone}} \bibnamefont{and}
  \bibinfo{author}{\bibfnamefont{G.}~\bibnamefont{Nisbet}},
  \bibinfo{journal}{New Surveys in the Classics} \textbf{\bibinfo{volume}{38}},
  \bibinfo{pages}{5} (\bibinfo{year}{2008}).

\bibitem[{\citenamefont{Semmann et~al.}(2003)\citenamefont{Semmann, Krambeck,
  and Milinski}}]{Semmann2003nature}
\bibinfo{author}{\bibfnamefont{D.}~\bibnamefont{Semmann}},
  \bibinfo{author}{\bibfnamefont{H.-J.} \bibnamefont{Krambeck}},
  \bibnamefont{and} \bibinfo{author}{\bibfnamefont{M.}~\bibnamefont{Milinski}},
  \bibinfo{journal}{Nature} \textbf{\bibinfo{volume}{425}},
  \bibinfo{pages}{390} (\bibinfo{year}{2003}).

\bibitem[{\citenamefont{Jukema and Piersma}(2006)}]{Jukema2006bl}
\bibinfo{author}{\bibfnamefont{J.}~\bibnamefont{Jukema}} \bibnamefont{and}
  \bibinfo{author}{\bibfnamefont{T.}~\bibnamefont{Piersma}},
  \bibinfo{journal}{Biol. Lett.} \textbf{\bibinfo{volume}{2}},
  \bibinfo{pages}{161} (\bibinfo{year}{2006}).

\bibitem[{\citenamefont{Sinervo and Lively}(1996)}]{Sinervo1996nature}
\bibinfo{author}{\bibfnamefont{B.}~\bibnamefont{Sinervo}} \bibnamefont{and}
  \bibinfo{author}{\bibfnamefont{C.~M.} \bibnamefont{Lively}},
  \bibinfo{journal}{Nature} \textbf{\bibinfo{volume}{380}},
  \bibinfo{pages}{240} (\bibinfo{year}{1996}).

\bibitem[{\citenamefont{Chippindale}(2013)}]{Chippindale2013me}
\bibinfo{author}{\bibfnamefont{A.~K.} \bibnamefont{Chippindale}},
  \bibinfo{journal}{Mol. Ecol.} \textbf{\bibinfo{volume}{22}},
  \bibinfo{pages}{1190} (\bibinfo{year}{2013}).

\bibitem[{\citenamefont{Shuster and Wade}(1991)}]{Shuster1991nature}
\bibinfo{author}{\bibfnamefont{S.~M.} \bibnamefont{Shuster}} \bibnamefont{and}
  \bibinfo{author}{\bibfnamefont{M.~J.} \bibnamefont{Wade}},
  \bibinfo{journal}{Nature} \textbf{\bibinfo{volume}{350}},
  \bibinfo{pages}{608} (\bibinfo{year}{1991}).

\bibitem[{\citenamefont{Kirkup and Riley}(2004)}]{Kirkup2004nature}
\bibinfo{author}{\bibfnamefont{B.~C.} \bibnamefont{Kirkup}} \bibnamefont{and}
  \bibinfo{author}{\bibfnamefont{M.~A.} \bibnamefont{Riley}},
  \bibinfo{journal}{Nature} \textbf{\bibinfo{volume}{428}},
  \bibinfo{pages}{412} (\bibinfo{year}{2004}).

\bibitem[{\citenamefont{Nash}(1951)}]{nash1951am}
\bibinfo{author}{\bibfnamefont{J.}~\bibnamefont{Nash}}, \bibinfo{journal}{Ann.
  Math.} pp. \bibinfo{pages}{286--295} (\bibinfo{year}{1951}).

\bibitem[{\citenamefont{Kalai and Lehrer}(1993)}]{Kalai1993eco}
\bibinfo{author}{\bibfnamefont{E.}~\bibnamefont{Kalai}} \bibnamefont{and}
  \bibinfo{author}{\bibfnamefont{E.}~\bibnamefont{Lehrer}},
  \bibinfo{journal}{Econometrica} \textbf{\bibinfo{volume}{61}},
  \bibinfo{pages}{1019} (\bibinfo{year}{1993}).

\bibitem[{\citenamefont{Holt and Roth}(2004)}]{holt2004pnas}
\bibinfo{author}{\bibfnamefont{C.~A.} \bibnamefont{Holt}} \bibnamefont{and}
  \bibinfo{author}{\bibfnamefont{A.~E.} \bibnamefont{Roth}},
  \bibinfo{journal}{Proc. Natl. Acad. Sci.} \textbf{\bibinfo{volume}{101}},
  \bibinfo{pages}{3999} (\bibinfo{year}{2004}).

\bibitem[{\citenamefont{Ianni}(2014)}]{Ianni2014JME}
\bibinfo{author}{\bibfnamefont{A.}~\bibnamefont{Ianni}}, \bibinfo{journal}{J.
  Math. Econ.} \textbf{\bibinfo{volume}{50}}, \bibinfo{pages}{148}
  (\bibinfo{year}{2014}).

\bibitem[{\citenamefont{Cressman and Tao}(2014)}]{Cressman2014pnas}
\bibinfo{author}{\bibfnamefont{R.}~\bibnamefont{Cressman}} \bibnamefont{and}
  \bibinfo{author}{\bibfnamefont{Y.}~\bibnamefont{Tao}},
  \bibinfo{journal}{Proc. Natl. Acad. Sci.} \textbf{\bibinfo{volume}{111}},
  \bibinfo{pages}{10810} (\bibinfo{year}{2014}).

\bibitem[{\citenamefont{Schuster et~al.}(1981)\citenamefont{Schuster, Sigmund,
  Hofbauer, Gottlieb, and Merz}}]{Schuster1981bc}
\bibinfo{author}{\bibfnamefont{P.}~\bibnamefont{Schuster}},
  \bibinfo{author}{\bibfnamefont{K.}~\bibnamefont{Sigmund}},
  \bibinfo{author}{\bibfnamefont{J.}~\bibnamefont{Hofbauer}},
  \bibinfo{author}{\bibfnamefont{R.}~\bibnamefont{Gottlieb}}, \bibnamefont{and}
  \bibinfo{author}{\bibfnamefont{P.}~\bibnamefont{Merz}},
  \bibinfo{journal}{Biol. Cybern.} \textbf{\bibinfo{volume}{40}},
  \bibinfo{pages}{17} (\bibinfo{year}{1981}).

\bibitem[{\citenamefont{Sato et~al.}(2002)\citenamefont{Sato, Akiyama, and
  Farmer}}]{Sato2002pnas}
\bibinfo{author}{\bibfnamefont{Y.}~\bibnamefont{Sato}},
  \bibinfo{author}{\bibfnamefont{E.}~\bibnamefont{Akiyama}}, \bibnamefont{and}
  \bibinfo{author}{\bibfnamefont{J.~D.} \bibnamefont{Farmer}},
  \bibinfo{journal}{Proc. Natl. Acad. Sci.} \textbf{\bibinfo{volume}{99}},
  \bibinfo{pages}{4748} (\bibinfo{year}{2002}).

\bibitem[{\citenamefont{Hofbauer and Huttegger}(2008)}]{Hofbauer2008JTB}
\bibinfo{author}{\bibfnamefont{J.}~\bibnamefont{Hofbauer}} \bibnamefont{and}
  \bibinfo{author}{\bibfnamefont{S.~M.} \bibnamefont{Huttegger}},
  \bibinfo{journal}{J. Theor. Biol.} \textbf{\bibinfo{volume}{254}},
  \bibinfo{pages}{843} (\bibinfo{year}{2008}).

\bibitem[{\citenamefont{Kleshnina et~al.}(2021)\citenamefont{Kleshnina,
  Streipert, Filar, and Chatterjee}}]{Kleshnina2021plos}
\bibinfo{author}{\bibfnamefont{M.}~\bibnamefont{Kleshnina}},
  \bibinfo{author}{\bibfnamefont{S.~S.} \bibnamefont{Streipert}},
  \bibinfo{author}{\bibfnamefont{J.~A.} \bibnamefont{Filar}}, \bibnamefont{and}
  \bibinfo{author}{\bibfnamefont{K.}~\bibnamefont{Chatterjee}},
  \bibinfo{journal}{PLoS Comput. Biol.} \textbf{\bibinfo{volume}{17}},
  \bibinfo{pages}{e1008523} (\bibinfo{year}{2021}).

\bibitem[{\citenamefont{Boyd}(1989)}]{Boyd1989jtb}
\bibinfo{author}{\bibfnamefont{R.}~\bibnamefont{Boyd}}, \bibinfo{journal}{J.
  Theor. Biol.} \textbf{\bibinfo{volume}{136}}, \bibinfo{pages}{47}
  (\bibinfo{year}{1989}).

\bibitem[{\citenamefont{Hopkins}(2002)}]{hopkins2002econometrica}
\bibinfo{author}{\bibfnamefont{E.}~\bibnamefont{Hopkins}},
  \bibinfo{journal}{Econometrica} \textbf{\bibinfo{volume}{70}},
  \bibinfo{pages}{2141} (\bibinfo{year}{2002}).

\bibitem[{\citenamefont{B{\"o}rgers and Sarin}(2000)}]{borgers2000IER}
\bibinfo{author}{\bibfnamefont{T.}~\bibnamefont{B{\"o}rgers}} \bibnamefont{and}
  \bibinfo{author}{\bibfnamefont{R.}~\bibnamefont{Sarin}},
  \bibinfo{journal}{Int. Econ. Rev.} \textbf{\bibinfo{volume}{41}},
  \bibinfo{pages}{921} (\bibinfo{year}{2000}).

\bibitem[{\citenamefont{Jablonka and Lamb}(2014)}]{jablonka2014mit}
\bibinfo{author}{\bibfnamefont{E.}~\bibnamefont{Jablonka}} \bibnamefont{and}
  \bibinfo{author}{\bibfnamefont{M.~J.} \bibnamefont{Lamb}},
  \emph{\bibinfo{title}{Evolution in four dimensions, revised edition: Genetic,
  epigenetic, behavioral, and symbolic variation in the history of life}}
  (\bibinfo{publisher}{MIT press, Cambridge}, \bibinfo{year}{2014}).

\bibitem[{\citenamefont{Kuznetsov}(2006)}]{Kuznetsov2006Scholarpedia}
\bibinfo{author}{\bibfnamefont{Y.}~\bibnamefont{Kuznetsov}},
  \bibinfo{journal}{Scholarpedia} \textbf{\bibinfo{volume}{1}},
  \bibinfo{pages}{1858} (\bibinfo{year}{2006}).

\bibitem[{\citenamefont{von Neumann and Morgenstern}(1944)}]{vNM}
\bibinfo{author}{\bibfnamefont{J.}~\bibnamefont{von Neumann}} \bibnamefont{and}
  \bibinfo{author}{\bibfnamefont{O.}~\bibnamefont{Morgenstern}},
  \emph{\bibinfo{title}{Theory of Games and Economic Behavior}}
  (\bibinfo{publisher}{Princeton University Press, Princeton},
  \bibinfo{year}{1944}).

\bibitem[{\citenamefont{Macy}(1991)}]{macy1991UCP}
\bibinfo{author}{\bibfnamefont{M.~W.} \bibnamefont{Macy}},
  \bibinfo{journal}{Am. J. Sociol.} \textbf{\bibinfo{volume}{97}},
  \bibinfo{pages}{808} (\bibinfo{year}{1991}).

\bibitem[{\citenamefont{Flache and Macy}(2002)}]{FLACHE2002JCR}
\bibinfo{author}{\bibfnamefont{A.}~\bibnamefont{Flache}} \bibnamefont{and}
  \bibinfo{author}{\bibfnamefont{M.~W.} \bibnamefont{Macy}},
  \bibinfo{journal}{J. Confl. Resolut.} \textbf{\bibinfo{volume}{46}},
  \bibinfo{pages}{629} (\bibinfo{year}{2002}).

\bibitem[{\citenamefont{Macy and Flache}(2002)}]{Macy2002PNAS}
\bibinfo{author}{\bibfnamefont{M.~W.} \bibnamefont{Macy}} \bibnamefont{and}
  \bibinfo{author}{\bibfnamefont{A.}~\bibnamefont{Flache}},
  \bibinfo{journal}{Proc. Natl. Acad. Sci.} \textbf{\bibinfo{volume}{99}},
  \bibinfo{pages}{7229} (\bibinfo{year}{2002}).

\end{thebibliography}

\end{document}